\documentclass[aps,prb,preprint,showpacs,superscriptaddress]{revtex4}
\usepackage{bm, amsmath, amssymb}
\usepackage{graphicx,subfigure}

\newcommand{\figwidth}{0.47\textwidth}

\begin{document}

\title{Collective modes for an array of magnetic dots with perpendicular
magnetization }

\author{P.~V. Bondarenko}
\affiliation{Institute of Magnetism, National Academy of Sciences of
Ukraine, 03142 Kiev, Ukraine}

\author{A.~Yu. Galkin}
\affiliation{Institute of Metal Physics, National Academy of
Sciences of Ukraine, 03142 Kiev, Ukraine }

\author{B.~A. Ivanov}
\email{bivanov@i.com.ua} \affiliation{Institute of Magnetism,
National Academy of Sciences of Ukraine, 03142 Kiev, Ukraine}
\affiliation{National Taras Shevchenko University of Kiev, 03127
Kiev, Ukraine}

\author{C.~E. Zaspel }
\affiliation{University of Montana-Western, Dillon, MT 59725, USA. }

\date{\today}

\begin{abstract}
The dispersion relations of collective oscillations of the magnetic
moment of magnetic dots arranged in square-planar arrays and having
magnetic moments perpendicular to the array plane are calculated.
The presence of the external magnetic field perpendicular to the
plane of array, as well as the uniaxial anisotropy for single dot
are taken into account. The ferromagnetic state with all the
magnetic moments parallel, and chessboard antiferromagnetic state
are considered. The dispersion relation yields information about the
stability of different states of the array. There is a critical
magnetic field below which the ferromagnetic state is unstable. The
antiferromagnetic state is stable for small enough magnetic fields.
The dispersion relation is non-analytic as the value of the wave
vector approaches zero. Non-trivial Van Hove anomalies are also
found for both ferromagnetic and antiferromagnetic states.
\end{abstract}

\maketitle

\section{Introduction}

During the last decade the most impressive achievements in magnetism
were related to fabrication, investigation, and application of
artificial magnetic materials (see Ref.~\onlinecite{Skomski} for a
recent review). The technologies of sputtering and lithography have
progressed to the state where the manufacture of nanosize, periodic
magnetic superlattices of different types is feasible. Among them
two-dimensional lattices of sub-micron magnetic particles (so-called
magnetic dots) attract much attention. These magnetic dots, in the
form of circular or elliptic cylinders, or rectangular prisms, are
made of soft magnetic materials such as Co and
permalloy,~\cite{Hillebrands,Miramond,Wassermann,Runge,Cowburn} or
highly anisotropic materials like Dy,~\cite{WeissKlitzing} and
FePt,~\cite{FePt} and the dot array lattice is usually designed to
be quadratic or rectangular. In the array dots are separated from
each other so that direct exchange interaction between dots is
completely absent. Thus the dipolar interaction is the sole source
of coupling between dots and determines the pattern of dot magnetic
moment orientations that constitutes the physical properties of a
dot array. Owing to the absence of exchange, magnetic dot arrays
constitute promising material for high-density magnetic storage
media. For this purposes, the dense arrays (with the period of order
of 100-200 nm) of small enough magnetic dots with the magnetic
moments perpendicular to the array plane are optimal, see Ref.
\onlinecite{Ross}. For small enough dots (the diameter $<$ 100 nm)
the magnetization inside of a dot is almost uniform, producing the
total magnetic moment $m_0 \gg\mu _B $, where $\mu _B $ is Bohr
magneton. Therefore, this can be a new kind of magnetic material
with purely two-dimensional lattice structure and pure dipolar
coupling between large enough magnetic moments.

To describe the physical properties of magnets in general it is essential to
take into account interactions having various origins and different energy
scales. For adjacent spins the spin exchange interaction is almost
invariably the strongest. If this interaction is ferromagnetic it causes
uniform spin ordering over distances that usually substantially exceeds the
atomic spacing. The dipole-dipole interaction, though as a rule is
considerably weaker than the exchange for adjacent spins, but it extends to
much longer range. The competition of these two interactions produces
magnetic domain structure with long-range non-uniformity of magnetization
for usual magnetic samples.

Models of magnetic moments with the dipolar interaction have been
theoretically studied for more than 60 years, and many physical
properties, lacking in the spin-exchanged systems, are known for
those models.\cite{LutTisza,BelobGIgnat} Note first the presence of
a non-unique ground state with non-trivial continuous degeneracy
that is typical both for three-dimensional lattices
\cite{BelobGIgnat} and for two-dimensional lattices of planar
\cite{DipD2,Gus99} and three-dimensional dipoles.\cite{Gus+00}
Two-dimensional systems of Ising-type dipoles with perpendicular
anisotropy demonstrate a complicated behavior when subject to an
external magnetic field perpendicular to the array's plane.~
\cite{BishGalkIv} At zero field, the ground state of a square 2D
array is chessboard antiferromagnetic (AFM), and this state is
stable at low fields $H<H_1 =2.3M$, where $M=m_0/a^3$ is
characteristic field of magnetic dipole interaction, and $a$ is the
lattice spacing. The ferromagnetic (FM) (saturated) state is stable
at high fields $H>H_0 =9.03 M$, with a cascade of phase transitions
between complicated magnetic structures at $H_1 < H < H_0$.

The models \cite{LutTisza,BelobGIgnat,DipD2} were discussed
originally in regard to the description of real crystalline spin
systems in which the dipolar interaction is predominant, such as
some rare-earth magnets. However, fabrication and experimental study
of magnetic dot arrays provide new physical systems for the testing
of basic magnetism models. For many materials, such as compounds
with rare-earth ions, granular magnets, and diluted solid solution
of paramagnetic ions in nonmagnetic crystals, the magnetic
properties seem to be similar to those of dot arrays. Nevertheless,
magnetic dot lattices exhibit physical properties, which are absent
in all above-mentioned systems. First, dot arrays in contrast to
layered crystals are literally two-dimensional. The manifestations
of long-ranged magnetic dipole interaction are principally different
for of layered crystals and truly two-dimensional systems.
\cite{Maleev+all} Second, the scale of dipolar interaction of two
spins $(2\mu _B S)^2/a^3$, where $\mu _B $ the Bohr magneton, $a$ is
the inter-spin distance, even with large spins like $S = 7/2$ does
not exceed several Kelvins, whereas for dots with volume as small as
10$^{4}$ nm$^{3}$ the characteristic magnetic moment $m_0 >10^3\mu
_B S$, and the characteristic energy can be comparable to or even
higher than the thermal energy at room temperature. Besides, for a
compound with high density of rare earth ions though the exchange
interaction is small it is not completely negligible.

As both FM and AFM exchange interactions of adjacent spins lead to
magnetic states essentially different from those caused by dipolar
interaction, these systems can not be considered as purely dipolar.
The large magnetic moment of each particle is typical also for
granular materials, but the magnetic dot lattices are different from
the latter by the high spatial regularity. It is worth mentioning a
new class of materials -- molecular crystals with high-spin
molecules.\cite{Wernsdorfer01} However, they are three-dimensional,
and the spin of these systems is much less than the effective spin
for magnetic dots not exceeding 10 or 15. It is also important that
the size of magneto-active part of a single high-spin molecule is
small compared to the total size of molecule; therefore the dipole
interaction is weak.

Hence, magnetic dot arrays are specific new magnetic materials with
purely two-dimensional and quite regular lattice structure and long
distance dipolar coupling between magnetic moments, which are rather
large and manifest at high temperature. From the point of view of
dynamical properties this implies that for magnetic dot arrays the
well-defined modes of collective oscillations characterized by
definite quasimomentum should exist, while this is admittedly not
the case for granular magnets or dilute solid solutions of
paramagnetic ions. The direct measurement of the dependence $\omega
(\vec {k})$ can be done by the Brillouin light scattering method. In
the pioneering experiments \cite{Mathieu97,Mathieu98,Jorzick99} no
indication for a band structure due to the periodic arrangement of
the dots was found, see also review
articles.~\cite{Demokr+Rev01,DemokrRev03} However, more recently the
dispersion effects for dense arrays were clearly observed by the
Brillouin light scattering method,~\cite{GubbiottiExper,Wang09Exp}
and time resolved scanning Kerr microscopy,~\cite{Kruglyak10}
stimulating the development of the theory for these systems.

In first theoretical articles, the finite systems were investigated,
sometimes with large enough number of dots $N$, such as $N\sim
10^3$-$10^4$, see Ref.~\onlinecite{ShibOtani} or even smaller
systems.~\cite{Giovannini3x3} The analytical calculations were
performed using the Bloch theorem which is applicable to infinite
arrays. The cases of spherical particles,
\cite{AriasMills04,AriasMills06,Tartak+08} cylindrical particles in
uniform state \cite{PolitiPini02,Galkin+VortJM3} and in the vortex
state,~\cite{Galkin+Vort} were also considered, see for
review.~\cite{Antos+Rev})

In the present work we considered collective modes for a square
lattice of rather small dots of nearly ellipsoidal shape in which
magnetization can be considered as uniform within a single dot in
the presence of an external bias magnetic field. Only the simplest
magnetic states of the magnetic dot system, namely, FM state and
chessboard AFM state are considered. The long-range character of
interaction of magnetic dots leads to unique properties of
collective excitations in this system, which are absent in both
continuous thin films and for dipole coupled spins in the 3D
lattice. Especially for the FM state it is clear that non-analytic
dependence of collective mode frequencies on quasi-momentum $\vec
{k}$, appears. Namely, as $\vec {k}\to 0 $ the spectrum has a finite
gap $\omega _0 $, but the dispersion law is nonanalytic $\omega \to
\omega _0 +c\vert \vec {k}\vert $. For AFM state, the spectrum
consists from two energy bands, which are connected at the border of
the Brillouin zone of the lattice at zero magnetic field, but these
two bands are well separated by an energy gap at finite fields. For
this state the unusual extremum, which is a saddle point in the
center of the Brillouin zone is found.

\section{ Model description. }
For uniform magnetization inside of each particle the state of the
dot array is described by the full magnetic moment of each dot,
$\vec {m}_{\vec {l}} $, $\left| {\vec {m}_{\vec {l}} } \right|=m_0 $
placed in the square lattice sites $\vec {l}=a(\vec {e}_x l_x +\vec
{e}_y l_y )$, where $l_x ,{\kern 1pt}\;l_y $ are integers. The
system Hamiltonian describing the dipole interaction of dots subject
to an external magnetic field $\vec {H}_0 $ and taking into account
the uniaxial magnetic anisotropy for each dot reads~\cite{SW}
\begin{eqnarray}
\nonumber \label{eq1}
  \hat {H}&=& \frac{1}{2}\sum\limits_{\vec {l}\ne {\vec
{l}}'} {\;\frac{\vec {m}_{\vec {l}} \vec {m}_{\vec {{l}'}} -3(\vec
{m}_{\vec {l}} \vec {\nu })(\vec {m}_{{\vec {l}}'} {\vec {\nu
}}')}{\left| {\vec {l}-{\vec {l}}'} \right|^3}} - \\
   &-& \sum\limits_{\vec {l}} {\left[ {\frac{1}{2}\kappa \cdot (\vec
{m}_{\vec{ l} } \vec {e}_z )^2+\vec {m}_{\vec{ l} } \vec {H}_0 }
\right]}\,.
\end{eqnarray}

Here $\vec {\nu }=(\vec {l}-{\vec {l}}')/\left| {\vec {l}-{\vec
{l}}'} \right|$ , $\kappa $ is the anisotropy constant for a given
dot, which is assumed to be uniaxial with a easy axis $\vec {e}_z $,
perpendicular to the dot array plane. For dots made of soft magnetic
material, the anisotropy is associated with the dot shape ($\kappa
>$ 0 or $\kappa  < 0$ for dots oblate or oblong along $\vec
{e}_z $, respectively). For dots made of highly anisotropic magnetic
materials,~\cite{WeissKlitzing,FePt} the crystalline anisotropy of
the material can provide some contribution to $\kappa $. The state
of a single dot can be characterized by perpendicular magnetization
($\vec {m}_z =\pm m_0 \vec {e}_z )$. However, in the following only
the most symmetric case will be considered where the external field
$\vec {H}_0 $ is perpendicular to the array plane.

Although oscillations of magnetization in the dot can be treated in
a purely classical manner, it is convenient to employ the operator
approach.~\cite{Galkin+Vort} For dots with homogeneous magnetization
it is natural to introduce magnon creation and annihilation
operators for the total magnetic moment. To find a dispersion law in
the linear approximation, it is sufficient to use the formulae,
\cite{SW}
\begin{eqnarray} \label{eq2} \nonumber
 m_{3,\vec {l}} &=&m_0 -g\mu _B a_{\vec {l}}^\dag a_{\vec {l}} , \\
 m_{1,\vec {l}} &=&(a_{\vec {l}}^\dag +a_{\vec {l}} ) \sqrt {gm_0 \mu _B /2} ,
 \\ \nonumber
 m_{2,\vec {l}} &=& i (a_{\vec {l}}^\dag -a_{\vec {l}}
 )\sqrt {gm_0 \mu _B/2 }\, ,
 \end{eqnarray}
below we will put Lande factor $g=2$, which in this approximation
are similar to the familiar Holstein-Primakoff or Dyson-Maleev
representations.~\cite{SW} Here $\mu _B$ is the Bohr magneton, 1, 2,
3 denote the projections on the set of orts, for example, $\vec
{e}_x $, $\vec {e}_y $, $\vec {e}_z $, for $\vec {m}=m_0 \vec {e}_z
$ or $\vec {e}_x $, $-\vec {e}_y $, $-\vec {e}_z $ for $\vec
{m}=-m_0 \vec {e}_z $.

\section{ Ferromagnetic state of array.}
For the FM state of the dot array ($\vec {m}=\vec {e}_z $ for all
dots) the same set of orts in Eq.~(\ref{eq2}), $\vec {e}_1 =\vec
{e}_x $, $\vec {e}_2 =\vec {e}_y $ and $\vec {e}_3 =\vec {e}_z $
should be used. In the quadratic approximation over the operators
$a_{\vec {l}}^\dag $ and $a_{\vec {l}} $ the Hamiltonian reads
\begin{widetext}
\begin{equation}\label{longHam}
 \hat{H}=2\mu_B\sum\limits_{\vec {l}} {\left[ {\left( {H_0 +H_a -\sum\limits_{\vec
{\delta }\ne 0} {\frac{M}{| {\vec {\delta }} |^3}} } \right)a_{\vec
{l}}^\dag a_{\vec {l}}
 -\frac{1}{2}\sum\limits_{\vec {\delta }\ne 0} {\frac{M}{| {\vec {\delta }} |^3}a_{\vec {l}}^\dag a_{\vec {l}+a\vec
{\delta }} } } \right]}
 -\frac{3\mu _B M}{2}\sum\limits_{\vec {l}} {\left[ {\sum\limits_{\vec
{\delta }\ne 0} {\frac{\left( {\delta _x +i\delta _y }
\right)^2}{\left| {\vec {\delta }} \right|^5}a_{\vec {l}}^\dag
a_{\vec {l}+a\vec {\delta }}^\dag +\mathrm{h.c.}} } \right]}
\end{equation}
\end{widetext}
where $\vec {\delta }=l_x \vec {e}_x +l_y \vec {e}_y $
is a dimensionless lattice vector, $H_a =\kappa m_0 $ is the
anisotropy field, $M=m_0 /a^3$ is the characteristic value defining
the dipolar interaction intensity and having the same dimension as
usual the 3D magnetization. The collective modes are introduced via
states $a_k $ and $a_k^\dag $ of definite quasi-momentum $\vec {k}$
\begin{equation}
\label{eq3} a_k =\frac{1}{\sqrt N }\sum\limits_{{\vec {l}}'}
{a_{\vec {l}} e^{i\vec {k}\vec {l}}} ,\;a_k^\dag =\frac{1}{\sqrt N
}\sum\limits_{\vec {l}} {a_{\vec {l}}^\dag e^{-i\vec {k}\vec {l}}},
\end{equation}
where $N$ is the total number of dots in an array. The collective
modes are defined by the quadratic Hamiltonian over $a_k $,
$a_k^\dag $ which acquires the standard form
\begin{equation}
\label{eq4} \hat{H}=2 \mu_B M \sum\limits_k {\left[ {A_k a_k^\dag
a_k +\frac{1}{2}\left( {B_k a_k^\dag a_{-k}^\dag +B_k^\ast a_k
a_{-k} } \right)} \right]} \, ,
\end{equation}

When the form of coefficients $A_{k}$ and $B_{k}$ are established,
the collective excitation energy $\varepsilon (\vec {k})=\hbar
\omega (\vec {k})$ may be found by means of $u-v$ Bogolyubov
transformations (see, for example, \cite{SW}) and universally reads
 $$\varepsilon (\vec {k})= 2\mu_B M \sqrt {A_k^2 -\vert
B_k \vert ^2}, \ \omega (\vec {k})=\gamma M \sqrt {A_k^2 -\vert B_k
\vert ^2},$$
 where $\gamma =2\mu _B /\hbar $ is the gyromagnetic ratio.
The concrete forms of $A_{k}$ and $B_{k}$ are defined by the
distribution of the magnetic moments within the array. For the case
of interest (parallel ordering of dot magnetization) one can find
\begin{equation}
\label{eq5} A_k=h +\beta -\frac{3}{2}\sigma (0)+\frac{1}{2}\left[
\sigma (0)-\sigma (\vec {k}) \right], \ B_k= 3 \sigma _c (\vec {k}),
\end{equation}
where $h=H_0/M$ and $\beta = H_a/M$ are dimensionless magnetic field
and anisotropy constant, respectively; and the dipolar sums $\sigma
(\vec {k})$ and $\sigma _c (\vec {k})$ appear
\begin{eqnarray}\nonumber
\label{eq6} \sigma (\vec {k})&=&\sum\limits_{\vec {l}\ne 0}
{\frac{1}{(l_x^2 +l_y^2 )^{3/2}}e^{i\vec {k}\vec {l}}}\,,  \\ \sigma
_c ( {\vec {k}})&=& {\sigma }'(\vec {k})+i{\sigma }''(\vec
{k})=\sum\limits_{\vec {l}\ne 0 } {\frac{(l_x -il_y )^2}{(l_x^2
+l_y^2 )^{3/2}} \cdot e^{i\vec {k}\vec {l}}}.
\end{eqnarray}

Such sums naturally emerge for any problem involving interaction of
dipoles ordered in a lattice. For the two-dimensional case the sums
$\sigma ( {\vec {k}})$ and $\sigma _c ( {\vec {k}})$ have a series
of peculiarities discussed in
Refs.~\onlinecite{PolitiPini02,Galkin+VortJM3,Galkin+Vort}. Now we
note only that the sum $\sigma ( {\vec {k}})$ at $\vec {k}=0$
converges for large $\vec {l}$ faster than in the three-dimensional
case and has the finite value $\sigma (0)=9.03362 $. On the other
hand, the representation $\sigma ( {\vec {k}})$ through the integral
applicable in the three-dimensional case (see Ref.~\onlinecite{SW})
is not feasible since the corresponding two-dimensional integral
$\int {dxdy/(x^2+y^2)^{3/2}} $  diverges as $x$, $y \to  0$.
Therefore, the regular presentation of the static demagnetization
field $\vec {H}_m $ through the magnetization components $\vec {M}$
of the form $\vec {H}_m =-4\pi \sum\limits_{i=1}^3 {\left( {\vec
{e}_i N_i M_i } \right)} $, where $N_{i}$ are demagnetizing factors,
$\sum\limits_{i=1}^3 {N_i } =1$, fails in the two-dimensional case.
In particular, Yafet and Georgy demonstrated~\cite{YafetGyorgy88}
that for the atomic monolayer the $z$-projection of static
demagnetizing field $\vec {H}_m $, is (3/2)$Ì\sigma (0)$ in our
notation, and it does not coincide with the continuum theory result
for a thin film, $\vec {H}_m =-4\pi \vec {e}_z M_z $. The
numerically obtained value of $\sigma (0)=9.03362$ gives $
(3/2)\sigma (0)=1.0783\cdot 4\pi $, thus the difference is nearly
8{\%}. For $\vec {k}\ne 0$ the specific feature of the
two-dimensional case manifests itself more vividly as the appearance
of a non-analytical dependence on $\vec {k}$, namely, at $k\to 0$,
\begin{equation}
\label{non-an} \sigma (\vec {k})=\sigma (0)-k\cdot F(\vec {k})
 , \sigma _c ( {\vec {k}})=\frac{(k_x -ik_y )^2}{\sqrt {k_x^2 +k_y^2 }
}\cdot G(\vec {k}),
\end{equation}
where $k=| {\vec {k}} |$, $G(\vec {k})$ and $F(\vec {k})$ are the
real functions, analytic as $k\to 0$ with the limit values $F(\vec
{k})\to 2\pi a$, $G(\vec {k})\to 2\pi a/3$ as $\vec {k}\to 0$ and it
is invariant under symmetry transformations of the square lattice.
Thus in the collective oscillation spectra non-analytic features
appear which are absent in the case of a three-dimensional
ferromagnet. From here it is seen that in contrast to $\sigma (
{\vec {k}})$ the complex sum $\sigma _c ( {\vec {k}})$ is not
invariant with respect to symmetry transformations of the square
lattice, but these transformations change only the phase factor, and
therefore do not have an effect on the frequency. The frequency
contains only $\left| {\sigma _c (\vec {k})} \right|=k\cdot G(\vec
{k})$,
\begin{multline}
\label{eq7} \omega (\vec {k})=\gamma \sqrt { H_a^{\mathrm{eff}} +H_0
+\frac{M}{2}\Sigma _{(+)} (\vec {k})} \\ \cdot \sqrt
{H_a^{\mathrm{eff}} +H_0 +\frac{M}{2} \Sigma _{(-)} (\vec {k})}  \,,
\end{multline}
where we introduced the combinations of dipole sums $\Sigma _{(\pm
)} (\vec {k})$
\begin{equation}
\label{eq8} \Sigma _{(\pm )} (\vec {k})=\sigma (0)-\sigma (\vec
{k})\pm 3\left| {\sigma _c (\vec {k})} \right|\,.
\end{equation}

These sums are plotted in Fig.~1 as a function of $\vec{k}$ for
symmetrical directions of the reciprocal lattice. The field,
$H_a^{\mathrm{eff}} =H_a -(3/2)M\sigma (0)$ is the $z$-projection of
the effective anisotropy field, comprising an anisotropy field for
the single particle and demagnetization field of the array as a
whole, and $\sigma \left( 0 \right)\left( {{3m_0 } \mathord{\left/
{\vphantom {{3m_0 } {2a^3}}} \right. \kern-\nulldelimiterspace}
{2a^3}} \right)\cong 1.08\cdot \left( {4\pi M} \right)$.

\begin{figure}[htbp]
\includegraphics[width=\figwidth]{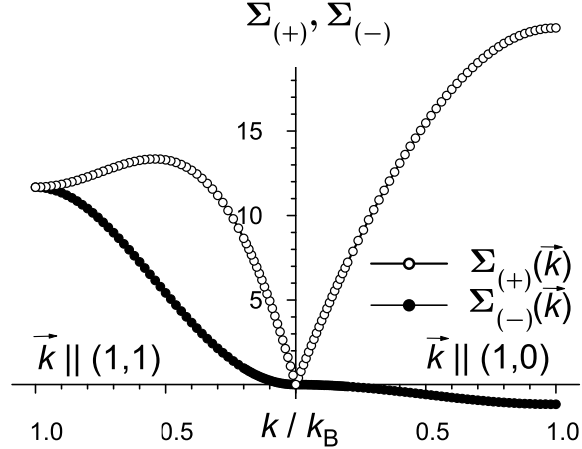}
\caption{\label{fig1}  The dependence of the combinations of the
dipole sums $\Sigma _{(\pm )} (\vec {k})$ on the quasimomentum for
symmetric directions of the square lattice, full symbols depict
$\Sigma _{(-)} (\vec {k})=\sigma (0)-\sigma (\vec {k})-3\left|
{\sigma _c (\vec {k})} \right|$, open circles represent $\Sigma
_{(+)} (\vec {k})=\sigma (0)-\sigma (\vec {k})+3\left| {\sigma _c
(\vec {k})} \right|$. $k_{B}$ is the maximal value of the wave
vector modulus for a given direction, corresponding to the border of
the Brillouin zone, $k_B =\pi /a$ for $\vec {k}\vert \vert
(0,1)\vert \vert $ (1,0) and $k_{B }=\pi $/$a $and $k_B =\pi \sqrt 2
/a$ for $\vec {k}\vert \vert (1,1)$. }
\end{figure}

\subsection{Dispersion relation and stability conditions. }
It is remarked that, the expression for the frequency can be
rewritten in the universal form
\begin{equation}
\label{eq9} \omega ^2(\vec {k})=\left[ {\omega _0 +\omega
_{\mathrm{int}} \Sigma _{(+)} (\vec {k})} \right]\left[ {\omega _0
+\omega _{\mathrm{int}} \Sigma _{(-)} (\vec {k})} \right],
\end{equation}
with the parameters $\omega _0 =\gamma (H_a^{\mathrm{eff}} +H_0 )$,
$\omega _{\mathrm{int}} =\omega _M LR^2/8a^3$, where $\omega _{0}$
can be thought of as the mode gap frequency, $\omega
_{\mathrm{int}}$ determines the magnitude of the mode dispersion
caused by interaction, $\omega _M =4\pi \gamma M_s $ is a
characteristic frequency of the material, which is $\omega _M $= 30
GHz for permalloy. It is worth noting, the same structure
(\ref{eq9}) appears for the collective mode of vortex precession for
the array of vortex state dots.~\cite{Galkin+Vort} To define the
parameters in (\ref{eq9}) we used $\omega _{\mathrm{int}} =\gamma
M/2=\gamma m_0 /2a^3$ and the value of the magnetic moment $m_0 =\pi
M_s LR^2$ for the single dot of cylindrical shape of thickness $L$
and radius $R$.

Let us discuss the characteristic features of the dispersion
relation represented by Eq. (\ref{eq9}) in more detail. For small
values of $\vert \vec {k}\vert $ both $\sigma (0)-\sigma (\vec {k})$
and $\left| {\sigma _c (\vec {k})} \right|$ are linearly increasing
functions of $| {\vec {k}} |$ for all directions of $\vec {k}$,
$\sigma (0)-\sigma (\vec {k})\to 2\pi a| {\vec {k}} |$, $\left|
{\sigma _c (\vec {k})} \right|\to 2\pi a| {\vec {k}} |/3$, as
explained in the Appendix. Because of this the first bracket in
Eq.~(\ref{eq9}) in the long-wave limit acquires the form
$H_{tot}^{\mathrm{eff}} +4\pi akM$, $k$ =$|\vec {k}|$. Note that the
multiplier before $k$, in contrast with that for static
demagnetizing field, is exactly the same as for continuum films. On
the other hand the linear components in $k$ compensate each other in
the second bracket as $\vec {k} \to 0$  and only the quadratic terms
in $k$ remain (see Fig. 1). Thus the magnon spectra have a
peculiarity as $\vec {k}\to 0$, $\omega ( {\vec {k}})\cong \omega _0
+2\pi \gamma Ma| {\vec {k}} |$ (see Fig. 2), while the value of
coefficient of $|\vec {k}|$ is exactly the same as for a thin
magnetic film of saturation magnetization, $M$ and thickness, $a$,
see Ref.~\onlinecite{GurevMelk,KalinikosSlavin}. Note the
interesting feature that the lattice constant of the array, related
to an in-plane space scale, plays the role of film thickness.

\begin{figure}[htbp]
\includegraphics[width=\figwidth]{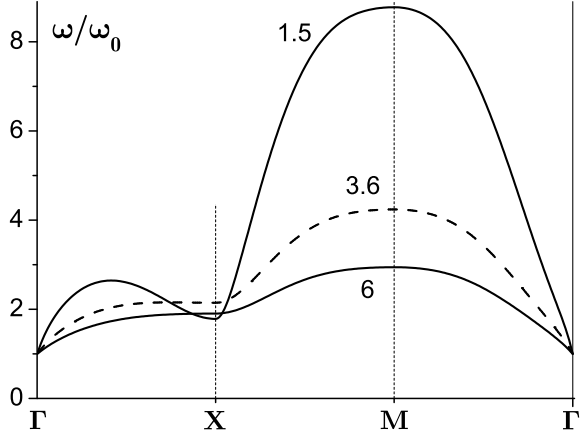}
\caption{\label{fig2}  The dispersion law for a dot array in the FM
state along some symmetric directions at different $\lambda =\omega
_0 /\omega _{\mathrm{int}} $ (shown near the curves). Here and below
we use the notations $\Gamma $, ${\rm X}$ and ${\rm M}$ for
symmetric points of the Brillouin zone (0,0), (0,1) and (1,1),
respectively. }
\end{figure}

Since the spectrum's behavior depends on the value of the external
magnetic field by a simple additive way, the sole parameter,
determining the form of the spectrum, is the ratio $\lambda =\omega
_0 /\omega _{\mathrm{int}} $. First, this dispersion relation is
strongly anisotropic. For all values of $\vert \vec {k}\vert $
inside the Brillouin zone the frequency $\omega (\vec {k})$ is
increasing monotonically for $\vec {k}$ parallel to the (1,1)
direction, but for $\vec {k}$ along the (1,0) axis the dependence
can be non-monotonic, as seen in Fig.~2. The oscillations have
$\omega ^{2} > 0$ and the ferromagnetic state is stable for weak
enough interaction, $\omega _0
>1.172\omega _{\mathrm{int}} $. Near the point of instability the
dependence $\omega ( {\vec {k}})$ for $\vec {k} \quad \vert \vert $
(1,0) has a minimum near the boundary of the Brillouin zone, which
is present until $\lambda =3.6$ which is also seen in Fig. 2. In all
this region of parameter space $1.172<\omega _0 /\omega
_{\mathrm{int}} <3.6$ some more extrema and saddle points are
present inside the Brillouin zone. For weaker interactions, $\omega
_{\mathrm{int}} <\omega _0 /3.6$, the dependence of $\omega ( {\vec
{k}})$ becomes monotonic inside all of the Brillouin zone.

Here it is remarked that the stability condition for the dot array
against small perturbations is nothing but the condition of positive
definiteness of the function $\omega ^{2}(\vec {k})$. The values of
the sums $\Sigma _{(\pm )} (\vec {k})$ vanishes as $k \to  0$, and
the stability of the FM state against the linear \textit{long wave}
excitations is broken if $H_a^{\mathrm{eff}} +H_0 <0$, or $H_{0} <
3\sigma (0)/2 - H_{a}$. After the replacement $3\sigma (0)/2\to 4\pi
$ and $M \to  M_{s}$ this criterion coincides with the result
obtained from the continuum theory for a thin magnetic film.
Nevertheless the dependence $\omega (\vec {k})$ is essentially
different than for thin films. The minimal value of the function
$\omega ^{2}(\vec {k})$ is attained at the boundary of the Brillouin
zone for the quasimomentum, $\vec {k}$ parallel to the (1,0) axis.
At this point, the maximal value of
 $$-(1/2)\Sigma _{(-)}(\vec
{k})=0.5[3\left| {\sigma _c (\vec {k})} \right|-\sigma (0)+\sigma
(\vec {k})]$$
 is equal to $0.5859=4\pi \cdot 0.0466$.
Because this instability condition is first realized for a non-small
$\vec {k}\parallel$ (1,0), it is stricter than the one found in the
long-wave approximation. Finally, the ferromagnetic state of the dot
array loses stability at $H_0 <H_c $, where
\begin{multline}
\label{eq10} H_c =\frac{M}{2}\max \left[ {3\left| {\sigma _c (\vec
{k})} \right|-\sigma (0)+\sigma (\vec {k})} \right]-H_a \simeq \\
\simeq 1.125\cdot 4\pi M-H_a .
\end{multline}
This value of the critical field differs from that obtained by the
continuum approximation, $4\pi M$ by more than 12~\%. Moreover, the
character of instability for the array of dots coupled by the dipole
interaction is principally different that for the continuous film.
For the dot array as $H_0 \le H_c $ the unstable mode has the
maximal increment at $\vec {k}=(\pi /a)\vec {e}_x $ or $\vec
{k}=(\pi /a)\vec {e}_y $ corresponding to the transition from the FM
state with parallel magnetic moments for all the dots to the
chessboard AFM state. For continuous films the instability occurs
only for long wave excitations and leads to appearance of
long-period domain structures.~\cite{BarIvJETP77}

\subsection{ Density of states. }
The complicated behavior of the dispersion curves play an important
role in the formation of Van Hove singularities. The Van-Hove
singularities are connected with the extrema of the dispersion law
of quasiparticles; namely, with minima, maxima and saddle points.
Any branch of collective excitations has at least one of these
extrema within the Brillouin zone. For the interaction of a finite
number of neighboring spins, the dispersion law is described by an
analytical function, and the function $\omega ( {\vec {k}})$ can be
approximated by parabolic functions in the vicinity of an extremum.
In this case the van Hove singularities have a standard form. In
particular, for a two-dimensional case the points of minima and
maxima of $\omega ( {\vec {k}})$, where $\omega =\omega _{\min } $
or $\omega =\omega _{\max } $, result in a finite jump of the
density of states, $D\left( \omega \right)=C\cdot \Theta \left(
{\omega -\omega _{\min } } \right)$ or $D\left( \omega
\right)=C\cdot \Theta \left( {\omega _{\max } -\omega } \right)$,
respectively, where $\Theta \left( x \right)$ is the Heaviside step
function, $\Theta \left( x \right)=1$ for $x >$ 0 and $\Theta \left(
x \right)=0$ for $x <0$. Saddle points with $\omega =\omega
_{\mathrm{c}} $ leads to logarithmic singularities of the form
$\Delta D\left( \omega \right)=C\cdot \ln \left[ {\omega
_{\mathrm{c}} /\left| {\omega -\omega _\mathrm{c} } \right|}
\right]$, and the appearance or disappearance of one more
singularity of such form at some value of frequency has to be
clearly seen.

In our case the structure can be richer due to the long-range character of
the interaction, as extrema can correspond to an non-standard behavior of
$D(\omega )$. First, note, the non-standard (linear in $k)$ dispersion near the
gap frequency, $\omega _0 $ produce much weaker singularities in the density
of states; for $\omega >\omega _0 $ simple calculations give
\begin{equation}
\label{eq11}
D(\omega )=C(\omega -\omega _0 )\cdot \Theta (\omega -\omega _0 )
\end{equation}
instead of finite jump. In addition, the number of extrema could be
more than three. In our case all these possibilities can be realized
at various values of the parameter $\lambda =\omega _0 /\omega
_{\mathrm{int}} $.

The analysis of the spectrum shows that for the FM state the
dependence $\omega =\omega (\vec {k})$ always has a minimum in the
point $\Gamma $ ($\vec {k}=0)$; moreover, in the vicinity of this
point $(\omega -\omega _0 )\propto | {\vec {k}} |$. As well $\omega
(\vec {k})$ always has the standard parabolic maximum at the point
${\rm M}$ ($\vec {k}=(1,1))$. Therefore, near the upper and lower
edges of the frequency band, the character of the singularities of
the density of states $D(\omega )$ is universal, $D(\omega )\propto
(\omega -\omega _0 )$ near the $\omega _0 $ (see Eq. (\ref{eq11})),
and $D(\omega )$ has a finite jump near the maximal frequency,
$\omega _{\max } $. At all values of $\lambda $, the logarithmic
singularity of the form of $\Delta D\left( \omega \right)=C\cdot \ln
\left[ {\omega _{\mathrm{c}} /\left| {\omega -\omega _{\mathrm{c}} }
\right|} \right]$ is also present (see Fig.~3).

\begin{figure}
  \subfigure[ ]{\includegraphics[width=\figwidth]{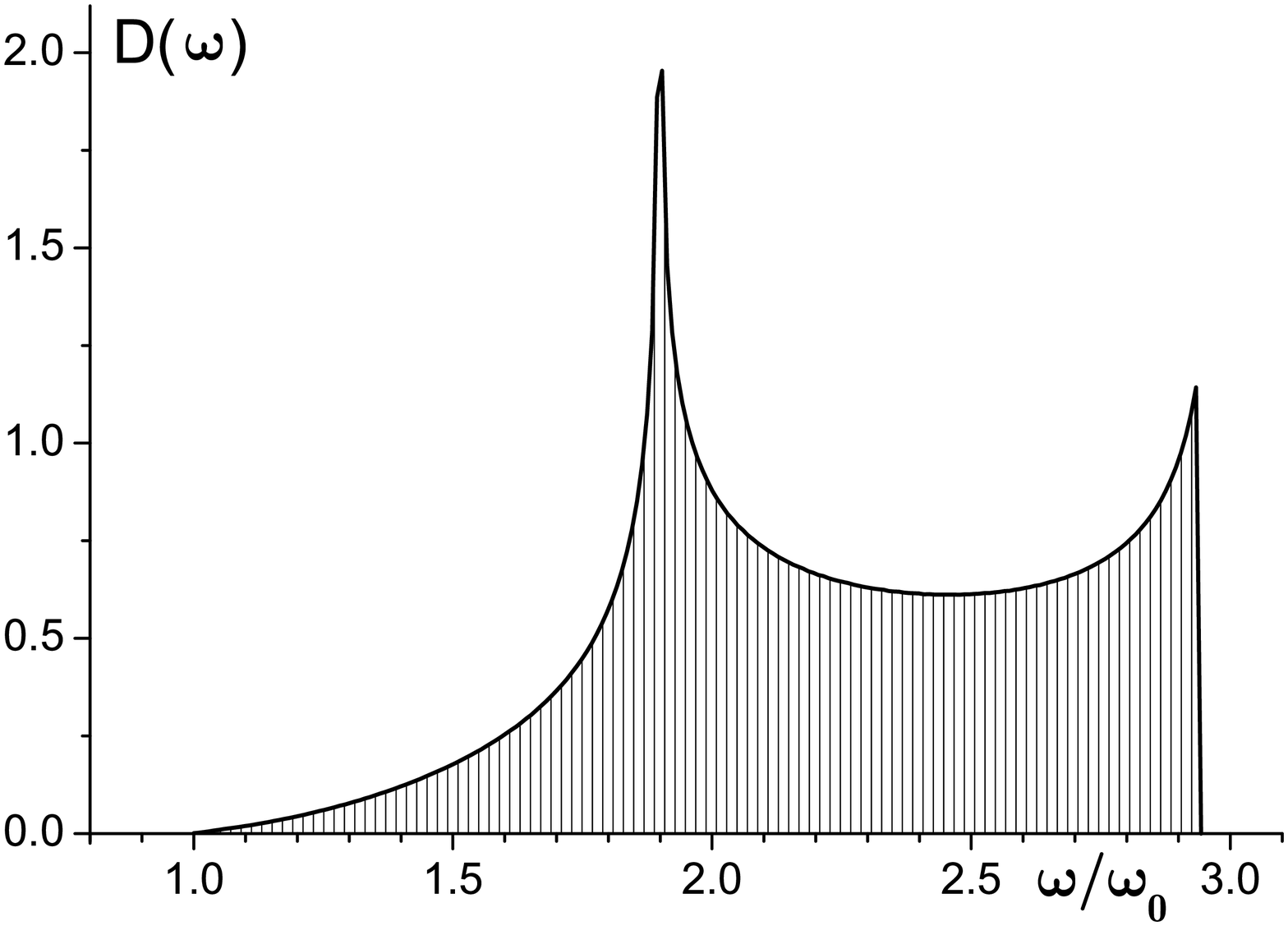}
    \label{exa1}}
    \subfigure[ ]{\includegraphics[width=\figwidth]{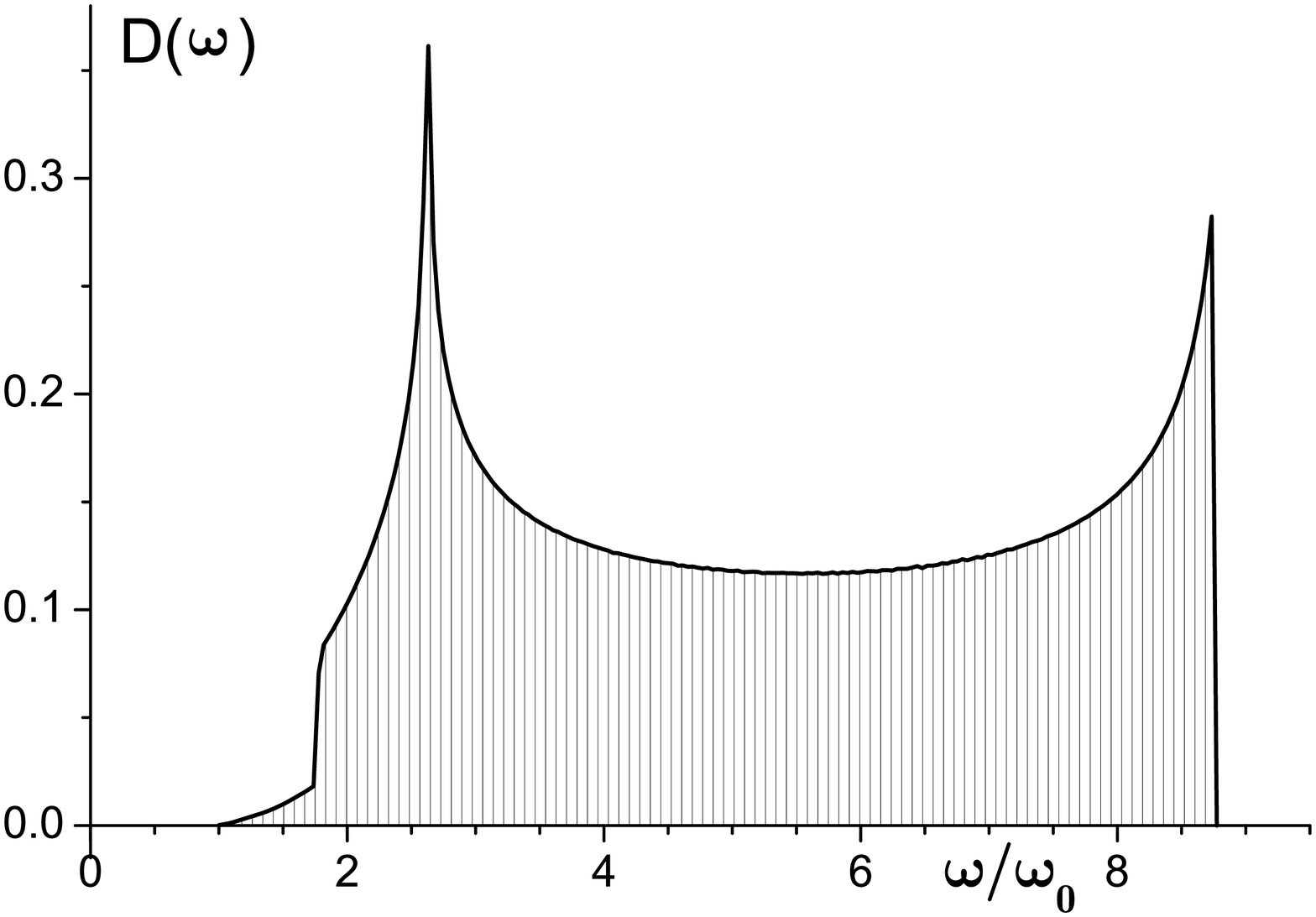}
    \label{exa1}}
    \subfigure[ ]{\includegraphics[width=\figwidth]{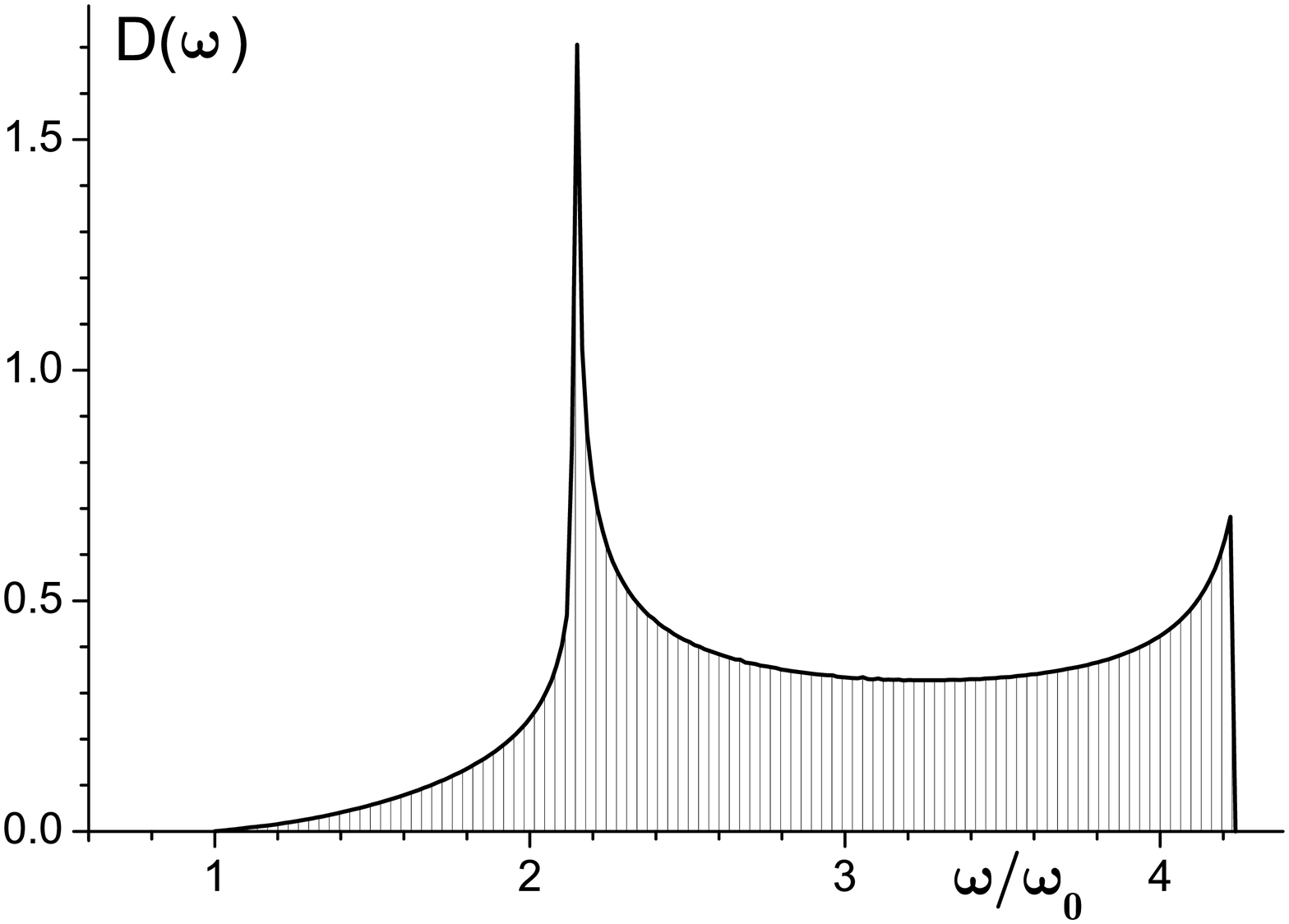}
    \label{exa1}}
       \caption{ The profile of the density of state function, normalized by
the condition $\int {D(\omega )d\omega } /\omega _0 =1$, for
different values of the parameter $\lambda =\omega _0 /\omega
_{\mathrm{int}} $. a) weak interaction, $\lambda =6>\lambda
_{\mathrm{c}} $; b) strong interaction, $\lambda =1.5<\lambda
_{\mathrm{c}} $; c) special case $\lambda =\lambda _{\mathrm{c}} $,
where more strong singularity appears. \label{fig3}}
\end{figure}

For large $\lambda $, corresponding to a weak interaction of
particles, the frequency grows with $\vert \vec {k}\vert $ for all
directions of $\vec {k}$. In this case the situation is standard:
saddle points are located at four symmetrical points of the type of
${\rm X}$ (1,0), and only the three aforementioned singularities are
present in the density of state function, see Fig.~3a. However, for
small $\lambda $, $\lambda <\lambda _{\mathrm{c}} =3.6$ the
dependence $\omega ( {\vec {k}})$ for $\vec {k}\vert \vert (0,1)$ is
non-monotonous: a local minimum with a standard parabolic dependence
is placed in the points of the type of ${\rm X}$ (1,0), where $\vec
{k}=\pm \pi \vec {e}_x /a$ or $\vec {k}=\pm \pi \vec {e}_y /a$. Thus
for small $\lambda $ four saddle points are symmetrically located
inside the Brillouin zone on the directions such as (0,1) with
$\vert \vec {k}\vert < \pi /a$. A local minimum leads to existence
of an additional jump of the density of states having the form
$D\left( \omega \right)=C_1 +C_2 \cdot \Theta \left( {\omega -\omega
_{lm} } \right)$ inside the frequency band for $\lambda <\lambda
_{\mathrm{c}} $, see Fig.~3b, whereas the saddle points produce
standard logarithmic singularities.

It is also seen that at the critical value of the parameter $\lambda
=\lambda _{\mathrm{c}} \simeq 3.6$, where two Van Hove singularities
emerge at some critical value, $\omega _{\mathrm{c}} $, the density
of states shows a new type of singularity
\begin{equation}
\label{eq12} D\left( \omega \right)=\frac{C}{\vert \omega -\omega
_{\mathrm{c}} \vert ^{1/4}}\, ,
\end{equation}
which is stronger than the standard logarithmic singularity (compare
Fig. 3c and Figs.~3a,b). To calculate this dependence, one have to
take into account the terms of order $(\vec {k}-\vec {k}_0 )^4$. It
is remarked here that the value of the parameter $\lambda =\omega _0
/\omega _{\mathrm{int}} $ depends on the external magnetic field and
can be changed continuously for the same sample during the
experiment, thus, the observation of this singularity is possible.
On the other hand, the non-standard (linear in $k)$ dispersion near
the gap frequency $\omega _{0}$ produce much weaker singularities in
the density of states. In particular, for $\omega  \ge  \omega _{0}$
simple calculations give $\Delta D\left( \omega \right)=C\cdot
(\omega -\omega _0 )\Theta \left( {\omega -\omega _0 } \right)$
instead of finite jump. This singularity is clearly seen for all
values of $\lambda $ (see Fig.3).

\section{ Chessboard antiferromagnetic state. }
In the previous section we investigated the states of the array in
which all dots are in the same magnetic state (FM ground state of
array). Next consider the AFM structure where the transition from FM
to AFM states can be controlled by a weak external magnetic field.
The spectral analysis of the dot array in the AFM state is of much
interest because this transition from FM to AFM states provides the
possibility to tune the properties of collective modes in the
system.

The method developed here might be easily generalized for the case of simple
AFM states, which might be described within the framework of a few
sublattices. To do that, it is necessary to split the array onto different
sublattices with the same dot state in each sublattice, and introduce
different Bose operators for each sublattice.

\subsection{ Brillouin zone, quasimomentum and magnon operators. }
Let us consider the simplest chessboard antiferromagnetic state,
with the magnetic moments $\vec {m}_{\vec l} =\left( {-1}
\right)^{l_x +l_y }m_0 \vec {e}_z $ in the ground state. This system
can be treated as the embedding of two quadratic sublattices with
the lattice spacing $a\sqrt 2 $, as shown in Fig. 4. Let us denote
the sites of the first sublattice which have $\vec {m}_{\vec {\delta
}} =\vec {e}_z $ as $a\vec {\delta }$, $\vec {\delta }=\sqrt 2
(\delta _1 \vec {\varepsilon }_1 +\delta _2 \vec {\varepsilon }_2
)$, where $\delta _1 $, $\delta _2 $ are integers and $\vec
{\varepsilon }_1 =(\vec {e}_x -\vec {e}_y )/\sqrt 2 $, $\vec
{\varepsilon }_2 =(\vec {e}_x +\vec {e}_y )/\sqrt 2 $ are basis
vectors for sublattices. Using orts $\vec {e}_x ,\;\vec {e}_y $, the
vector $\vec {\delta }$ can be expressed as $\vec {\delta }=\delta
_x \vec {e}_x +\delta _y \vec {e}_y $, where $\delta _x =\delta _1
+\delta _2 $, $\delta _y =\delta _2 -\delta _1 $. The sites for
second sublattice which have $\vec {m}_{\vec {\mu }} =-m_0 \vec
{e}_z $ can be expressed as $a\vec {\mu }$, $\vec {\mu }=\sqrt 2
[(\mu _1 +1/2)\vec {\varepsilon }_1 +(\mu _2 +1/2)\vec {\varepsilon
}_2 ]$, or $\vec {\mu }=\mu _x \vec {e}_x +\mu _y \vec {e}_y $, $\mu
_x =\mu _1 +\mu _2 +1$, $\mu _y =\mu _2 -\mu _1 $, where $\mu _1 $
and $\mu _2 $ are integers.

\begin{figure}[htbp]
\includegraphics[width=\figwidth]{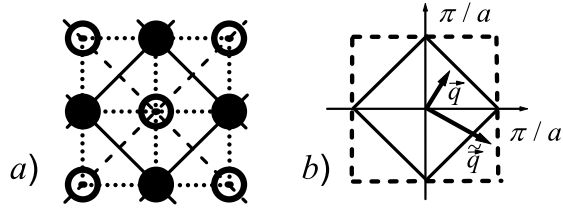}
\caption{\label{fig1} à) The scheme of the sublattices for the
particle array with the chessboard magnetic order; the solid and
dashed lines connect particles belonging to the first and second
sublattices, correspondingly; dotted lines connect particles from
different sublattices. b) the shape of the first Brillouin zone, the
solid line restricts the Brillouin zone of the sublattice, the
dashed line restricts the Brillouin zone for the whole lattice.  The
correspondence between  $\vec{q}$ and  $\tilde{\vec{q}}$, introduced
below, is also presented. }
\end{figure}

To describe the magnetic oscillations for two sublattices, it is
sufficient to use two Bose operators as follows,
\begin{eqnarray}\nonumber
m_{z,\vec {\delta }} &=&m_0 -2\mu _B a_{\vec {\delta }}^\dag a_{\vec
{\delta }} \,, m_{x,\vec {\delta }} =\left( {a_{\vec {\delta }}
+a_{\vec {\delta }}^\dag } \right)\sqrt {m_0 \mu _B } , \\
m_{y,\vec {\delta }} &=&i\left( {a_{\vec {\delta }}^\dag -a_{\vec
{\delta }} } \right)\sqrt {m_0 \mu _B } ; \, m_{z,\vec {\mu }} =2\mu
_B b_{\vec {\mu }}^\dag b_{\vec {\mu }} -m_0 , \\ \nonumber
m_{x,\vec {\mu }} &=&\left( {b_{\vec {\mu }} +b_{\vec {\mu }}^\dag }
\right)\sqrt {m_0 \mu _B } , \, m_{y,\vec {\mu }} =-i\left( {b_{\vec
{\mu }}^\dag -b_{\vec {\mu }} } \right)\sqrt {m_0 \mu _B } \,.
\end{eqnarray}

Then the Hamiltonian  takes the form, containing the terms
describing the interaction of oscillations both within first and
second sublattices, as well as between different sublattices. The
form of the corresponding terms is common to that for FM state, see
Eq.~\eqref{longHam}, and we do not present this long expression
here.

The translation symmetry within any sublattice allows us to introduce magnon
creation and annihilation operators for magnetic oscillations through the
Bose operators for each of the sublattices
\begin{equation}
\label{eq15}
a_{\vec {\delta }} =\frac{1}{\sqrt N }\sum\limits_{\vec {q}} {a_{\vec {q}}
e^{ia\vec {q}\vec {\delta }}} ,
\quad
b_{\vec {\mu }} =\frac{1}{\sqrt N }\sum\limits_{\vec {q}} {b_{\vec {q}}
e^{ia\vec {q}\vec {\mu }}} ,
\end{equation}
where, naturally, the quasi-momentum $\vec {q}$ takes values within
the first Brillouin zone of the sublattice (see Fig.~4). In the
basis of the sublattices the vector $\vec {q}=q_1 \vec {\varepsilon
}_1 +q_2 \vec {\varepsilon }_2 $, where $\left| {q_{1,2} }
\right|\le \pi /a\sqrt 2 $, and in the basis of the basic lattice
the vector $\vec {q}=q_x \vec {e}_x +q_y \vec {e}_y $, where $q_x
=(q_1 +q_2 )/\sqrt 2 $, $q_y =q_x =(q_2 -q_1 )/\sqrt 2 $, $\left|
{q_{x,y} } \right|\le \pi /a$. After this transform, the Hamiltonian
acquires the standard form $\hat H=\sum\limits_{\vec {q}} {\hat
H_{\vec {q}} } $, where
\begin{multline}
\label{eq16}
 \frac{\hat H_{\vec {q}}}{2\mu_B M} =(A_{\vec {q}} +h)a_{\vec {q}}^\dag a_{\vec {q}} +(A_{\vec
{q}} -h)b_{\vec {q}}^\dag b_{\vec {q}} - \\
 -\left[ {\frac{C_{\vec {q}}}{2} \left( {a_{\vec {q}}^\dag a_{-\vec
{q}}^\dag +b_{\vec {q}} b_{-\vec {q}} } \right)+D_{\vec {q}} a_{\vec
{q}}^\dag b_{\vec {q}} +F_{\vec {q}} a_{\vec {q}} b_{-\vec {q}}
+h.c.} \right],
\end{multline}
where the following notation are used
\begin{multline}
\label{eq17} A_{\vec {q}} =\sum\limits_{\vec {\mu }} {\frac{1}{|
{\vec {\mu }} |^3}-} \sum\limits_{\vec {\delta }\ne 0} {\frac{1}{|
{\vec {\delta }} |^3}} -\frac{1}{2}\sum\limits_{\vec {\delta }\ne 0}
{\frac{e^{ia\vec {q}\vec {\delta }}}{| {\vec {\delta }} |^3}} +\beta
\,,
\\ C_{\vec {q}} =\frac{3}{2}\sum\limits_{\vec {\delta }\ne 0}
{\frac{\left( {\delta _x +i\delta _y } \right)^2e^{ia\vec {q}\vec
{\delta }}}{| {\vec {\delta }} |^5}}\, ,\\ D_{\vec {q}}
=\frac{3}{2}\sum\limits_{\vec {\mu }} {\frac{\left( {\mu _x +i\mu _y
} \right)^2e^{ia\vec {q}\vec {\mu }}}{\left| {\vec {\mu }}
\right|^5}}\,, \ F_{\vec {q}} =\frac{1}{2}\sum\limits_{\vec {\mu }}
{\frac{e^{ia\vec {q}\vec {\mu }}}{| {\vec {\mu }} |^3}} .
\end{multline}
There is a simple connection between the sums over the sublattices
and the previously introduced sums over the whole lattice $\sigma
(\vec {k})$ and $\sigma _{\mathrm{c}} (\vec {k})$, see Eq.
\eqref{non-an}. For example, a simple geometrical transformation
gives
$$
\sum\limits_{\vec {\delta }\ne 0} {\frac{e^{ia\vec {q}\vec {\delta
}}}{| {\vec {\delta }} |^3}=} \frac{1}{2^{3/2}}\sigma \left( {\tilde
{\vec {q}}} \right),
$$
where we introduced the vector $\tilde {\vec {q}}=\sqrt 2 (q_1 \vec
{e}_x +q_2 \vec {e}_y )$, which is derived from the vector $\vec
{q}$ by rotation by the angle $\pi /4$ and by stretching by the
value $\sqrt 2 $ (see Fig. 4). It is easy to see that for the $\vec
{q}$ values within the first Brillouin zone of the sublattice, the
corresponding values of $\tilde {\vec {q}}$ are within the first
Brillouin zone of the whole lattice. By using the rules of
transition between the sums over the sublattices and the sum over
the whole lattice one can obtain
\begin{multline}
\label{eq17} A_{\vec {q}} =\left( {1-\frac{1}{\sqrt 2 }}
\right)\sigma \left( 0 \right)-\frac{1}{2^{5/2}}\sigma \left(
{\tilde {\vec {q}}} \right)+\beta ,\\ C_{\vec {q}}
=-\frac{3i}{2^{5/2}}\sigma _{\mathrm{c}} (\tilde {\vec {q}})\,,\
D_{\vec {q}} =\frac{3}{2}\left( {\sigma _{\mathrm{c}} (\vec
{q})+\frac{i}{2^{3/2}}\sigma _{\mathrm{c}} \left( {\tilde {\vec
{q}}} \right)} \right), \\ F_{\vec {q}} =\frac{1}{2}\left[ {\sigma
\left( {\vec {q}} \right)-\frac{1}{2^{3/2}}\sigma \left( {\tilde
{\vec {q}}} \right)} \right]\,.
\end{multline}
Thus, the Hamiltonian coefficients describing small oscillations of
the AFM state do not include new dipole sums different from those
for the FM case and could be expressed through the sums by means of
cumbersome, but simple geometrical transformations. It is necessary
to note that the Hamiltonian coefficients $F_{\vec {q}} $ and
$D_{\vec {q}} $ in the terms like $a_{\vec {q}} b_{-\vec {q}} $,
$a_{\vec {q}} b_{\vec {q}}^\dag $, responsible for interaction
between sublattices, are periodic relative to vectors of the
reciprocal lattice of the whole system, i.e. they have lower
symmetry than the coefficients responsible for interaction within
the sublattice. However, it can be proofed that in final expressions
for collective mode frequencies the translation symmetry of the
reciprocal lattice of period $\pi /a\sqrt 2 $ is restored.

\subsection{ Dispersion relation. }
For diagonalization of the Hamiltonian \eqref{eq16} one can use the
generalized Bogolyubov $u-v$ transformation and introduce the
creation and annihilation operators of magnons of two branches,
$c_{\vec {q}}^\dag $, $c_{\vec {q}} $ and $d_{\vec {q}}^\dag $,
$d_{\vec {q}} $, and the creation and annihilation operators of
different branches commute, $[c_{\vec {q}} ,d_{\vec {q}} ]=0$,
$[c_{\vec {q}} ,d_{\vec {q}}^\dag ]=0$. For normal modes $\dot
{c}_{\vec {q}} =-i\omega _{(-)} (\vec {q})c_{\vec {q}} $, $\dot
{d}_{\vec {q}} =-i\omega _{(+)} (\vec {q})d_{\vec {q}} $, where
$\omega _{(-)} (\vec {q})$ and $\omega _{(+)} (\vec {q})$ are the
frequencies of magnon modes. The generalized Bogolyubov transform
can be written as
\begin{multline}
\label{eq17} a_{\vec {q}} =u_{\vec {q}} c_{\vec {q}} +v_{\vec
{q}}^\ast c_{-\vec {q}}^\dag +{u}'_{\vec {q}} d_{\vec {q}} +v_{\vec
{q}}^{'\ast} d_{-\vec {q}}^\dag \, , \\ b_{\vec {q}} =\xi _{\vec
{q}} c_{\vec {q}} +\eta _{\vec {q}}^\ast c_{-\vec {q}}^\dag +{\xi
}'_{\vec {q}} d_{\vec {q}} +\eta _{\vec {q}}^{'\ast} d_{-\vec
{q}}^\dag \,.
\end{multline}

Comparing the equations of motion for the operators $c_{\vec {q}} $,
$d_{\vec {q}} $ (for example, $\dot {c}_{\vec {q}} =-i\omega _{(-)}
(\vec {q})c_{\vec {q}} )$ and the operators $a_{\vec {q}} $,
$b_{\vec {q}} $ ($i\hbar \dot {a}_{\vec {q}} =[a_{\vec {q}},\, \hat
H])$, the system of equations for the coefficients is presented as a
unitary transformation:
\begin{widetext}
\begin{equation}
\label{eq18}
\omega _{(\pm )} \left( {\vec {q}} \right)\left( {{\begin{array}{*{20}c}
 {u_{\vec {q}} } \hfill \\
 {v_{\vec {q}} } \hfill \\
 {\xi _{\vec {q}} } \hfill \\
 {\eta _{\vec {q}} } \hfill \\
\end{array} }} \right)=\left( {{\begin{array}{*{20}c}
 {(A_{\vec {q}} +h)} \hfill & {-C_{\vec {q}} } \hfill & {-D_{\vec {q}} }
\hfill & {-F_{\vec {q}} } \hfill \\
 {C_{\vec {q}}^\ast } \hfill & {-(A_{\vec {q}} +h)} \hfill & {F_{\vec {q}} }
\hfill & {D_{\vec {q}}^\ast } \hfill \\
 {-D_{\vec {q}}^\ast } \hfill & {-F_{\vec {q}} } \hfill & {(A_{\vec {q}}
-h)} \hfill & {-C_{\vec {q}}^\ast } \hfill \\
 {F_{\vec {q}} } \hfill & {D_{\vec {q}} } \hfill & {C_{\vec {q}} } \hfill &
{-(A_{\vec {q}} -h)} \hfill \\
\end{array} }} \right)\left( {{\begin{array}{*{20}c}
 {u_{\vec {q}} } \hfill \\
 {v_{\vec {q}} } \hfill \\
 {\xi _{\vec {q}} } \hfill \\
 {\eta _{\vec {q}} } \hfill \\
\end{array} }} \right)\, .
\end{equation}
\end{widetext}
In accordance with (\ref{eq18}) the frequencies of
collective modes of oscillations of dot magnetic moments, are
defined by a rather cumbersome formulae
\begin{equation}
\label{eq19}
\begin{array}{l}
 \omega _{(\pm )} ^2 (\vec q)=\left( {\gamma M} \right)^2\left[
{A^2+\left| D \right|^2+h^2-F^2-\left| C \right|^2} \right. \\
 \left. {\pm 2\sqrt {h^2\left( {A^2-F^2} \right)+\left| {AD+FC}
\right|^2-\left[ {\mathrm{Im}(CD^\ast) } \right]^2} } \right] \\
 \end{array}\,
\end{equation}
but an analysis of the dispersion relation can be done numerically.
The dependence of $\omega \left( {\vec {q}} \right)$ for the two
branches, $\omega _{(-)} \left( {\vec {q}} \right)$ and $\omega
_{(+)} \left( {\vec {q}} \right)$, for specific values of a magnetic
field are presented in Fig.~5.

\begin{figure}[htbp]
\includegraphics[width=\figwidth]{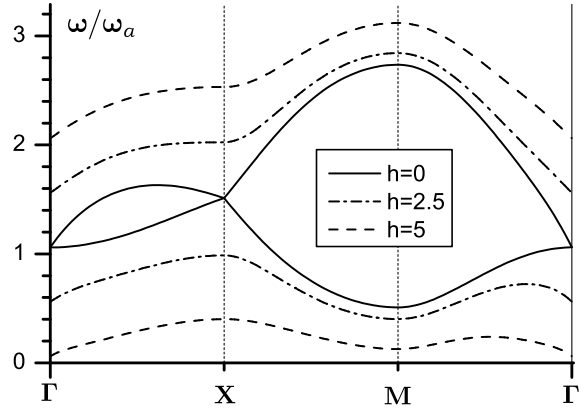}
\caption{\label{fig5}  The dispersion law along some symmetric
directions for a dot array with a moderate anisotropy value ($\beta
=5)$ in the AFM state at different values of the magnetic field $h$
(shown on the figure). The notations $\Gamma $, ${\rm X}$ and ${\rm
M}$ are the same as on Fig. 2, but for the Brillouin zone for
sublattice. }
\end{figure}

The given expression is essentially simplified at symmetrical points
of the Brillouin zone (in the zone center and on the edges), where
the expression
$$
\mathrm{Im} (CD^\ast) =\frac{9}{2^{7/2}}Re\left( {\sigma
_{\mathrm{c}}^\ast (\vec {q})\sigma _{\mathrm{c}} (\tilde {\vec
{q}})} \right)
$$
becomes zero at these symmetrical point.

The spectrum behavior essentially depends on the external magnetic
field value. The important and quite interesting (see below) case of
small fields is determined by the behavior at $h=0$. Therefore, we
begin with the study of asymptotics of the spectrum in the absence
of an external magnetic field. Because of the long-range nature of
the dipole interaction, the magnon spectrum has specific
peculiarities in the center of the Brillouin zone, therefore it is
reasonable to start with an analysis of the long wave limit $k\to
0$.

To first order in $| {\vec {k}} |$ the sums take the form (7) and
after simple algebra one obtains the asymptotic form of the
dispersion relation at $h=0$,
\begin{multline}
\label{eq20} \omega _\pm \left( {\vec {q}} \right)=\gamma M\sqrt {
{\beta +\left( {3\sqrt 2 /4} \right)\left( {\sqrt 2 -1}
\right)\sigma \left( 0 \right)} }
\\ \cdot \sqrt
{ {\beta -(1/2)\left( {\sqrt 2 -1} \right)\sigma \left( 0
\right)+\pi a\left| {\vec {q}} \right|\left( {1\pm 1} \right)} }\, .
\end{multline}
This expression describes a linear growth of frequency of the upper
branch with increasing $q\equiv \vert \vec {q}\vert $ in the
vicinity of the center of Brillouin zone. On the other hand, for the
frequency of the lower branch $\omega _- \left( {\vec {q}} \right)$
the terms linear in $\vec {q}$ are canceled out, and one can expect
that for this branch the dependence on the components of the vector
$\vec {q}$ should be parabolic. The numerical data presented in
Fig.~5 indicate this. However, there is also a sharp anisotropy of
the dependence $\omega _- \left( {\vec {q}} \right)$; namely, in the
vicinity of the point $\vec {q}=0$ the frequency $\omega _- \left(
{\vec {q}} \right)$ increases with a $q=\vert \vec {q}\vert $, if
$\vec {q}$ is parallel to one of directions (1,0), and $\omega _-
\left( {\vec {q}} \right)$ decreases if $\vec {q}$ is parallel to
(1,1). Note by virtue of forth order symmetry of the point $\vec
{q}=0$ such behavior is forbidden for the quadratic terms. The
anisotropy in the expansion of any analytical function $\omega (\vec
{q})$ in the $\vec {q}$ components can appear only due to invariants
of the fourth order, like $(q_x^4 +q_y^4 )$ or $q_x^2 q_y^2 $.
Actually, again a non-analyticity, caused by a slow convergence of
the dipolar sums, takes place here. Our analysis shows, in the
vicinity of the point $\vec {q}=0$ the dispersion relation can be
approximated by the expression
\begin{equation}
\label{eq21} \omega (\vec {q})\approx \omega _0 +\frac{\alpha (q_x^4
+q_y^4 )-2\beta q_x^2 q_y^2 }{\vert \vec {q}\vert ^2}\, ,
\end{equation}
with the values of the coefficients, $\alpha \simeq $0.654 and
$\beta \simeq 1.8$. Thus, in the zero field case a new type of
singular behavior of $\omega (\vec {q})$, namely, the presence of
specific saddle point forth-fold symmetry, appears in the lower
magnon branch.

In the case of nonzero external magnetic field in the vicinity of
the point $q\to 0$ one gets the following expression at $\left|
{\vec {q}} \right|\ll h$ accurate to within first order in $\left|
{\vec {q}} \right|$
\begin{multline}
\label{eq22} \omega _\pm \left( {\vec {q}} \right)=\gamma M\sqrt {
{\beta +(3\sqrt 2 /4)\left( {\sqrt 2 -1} \right)\sigma \left( 0
\right)} } \\ \cdot \sqrt { {\beta -(1/2)\left( {\sqrt 2 -1}
\right)\sigma \left( 0 \right)+\pi a\left| {\vec {q}} \right|} } \pm
\gamma H.
\end{multline}
The Eq. (\ref{eq22}) shows that the frequencies of both branches of
the spectrum in the vicinity of the center of the Brillouin zone
increase linearly with $| {\vec {q}} |$ at a nonzero magnetic field.

\subsection{ Stability of AFM state. }
A study of the obtained dispersion relation allows a determination
of a stability region for the AFM state, i.e. a region of parameters
at which $\omega ^2(\vec {q})$ becomes negative. The problem of
finding the stability region for the AFM state is essentially
simplified by the fact that the local minima of the lower branch of
the magnon spectrum are located in the center ($\vec {q}_{(0,0)}
=0)$ or in the corners ($\vec {q}_{(1,1)} =\pi (\vec {\varepsilon
}_1 +\vec {\varepsilon }_2 )/(a\sqrt 2 ))$ of the Brillouin zone of
the sublattice, see Fig.~5. At these points some coefficients of the
Hamiltonian vanish, that makes a simple analytical consideration
possible.

According to numerical calculations, in the absence of the external
magnetic field and at small fields, the minimal value of frequency
corresponds to magnons having the maximal value of quasimomentum,
$\vec {q}_{(1,1)} =\pi (\vec {\varepsilon }_1 +\vec {\varepsilon }_2
)/(a\sqrt 2 )=\pi \vec {e}_x /a$. Also at this point $\omega _\pm
(1,1)=A_{(1,1)} \pm \sqrt {D_{(1,1)}^2 +h^2} $. An analysis of
$\omega ^2(\vec {q})$ in the vicinity of this point gives that at
zero field the AFM state is stable only for enough large value of
anisotropy, $\beta >\beta _{\mathrm{cr}} $,
\begin{equation}
\label{eq23} \beta _{\mathrm{cr}} =-\frac{3}{2}\sigma _{\mathrm{c}}
\left( {\pi /a,0} \right)-\frac{7-3\sqrt 2 }{8}\sigma \left( 0
\right)\approx 2.4532.
\end{equation}
If $\beta >\beta _{\mbox{cr}} $, then with increasing of the field
the AFM state loses its stability relative to perturbations with
$\vec {q}_{(1,1)} $ at $\left| h \right|>h_{\mathrm{cr},11} =\sqrt
{A_{(1,1)}^2 -D_{(1,1)}^2 } $, see. Fig. 5.

\begin{figure}[htbp]
\includegraphics[width=\figwidth]{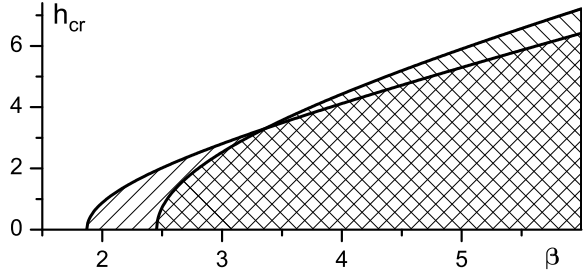}
\caption{\label{fig6}  The dependence of the critical magnetic field
on anisotropy. Here both characteristic fields, $h_{\mbox{cr},0} $
and $h_{\mathrm{cr},11} $ are plotted as functions of $\beta $, and
the stability region of the AFM state is doubly patterned region
under both curves. }
\end{figure}

The analysis also shows that for $\left| h \right|>h_{\mathrm{cr},0}
=\sqrt {A_0^2 -F_0^2 } $ the AFM state loses the stability relative
to small $\vec {q}$ perturbations. As it is seen from Fig. 5, both
these conditions are important at different values of anisotropy.
Thus, the AFM state is stable relative to arbitrary small
perturbations subject to the condition, $h<h_{\mathrm{cr}}
=\mathrm{min}\{h_{\mathrm{cr},0} ,\;h_{\mathrm{cr},11} \}$. For
small anisotropy $\beta <\beta _1 =3.358$ an instability with
maximum values of $q$ is developed, whereas at $\beta >\beta _1 $
the AFM state is unstable relative to long-wave perturbations with
$q\ll 1/a$.

\subsection{ Density of states. }
Let us consider positions and form of Van Hove singularities for the
spectral density of magnons of both branches as shown in Fig.~7. In
a weak field, frequency bands corresponding to the upper and lower
branches of collective oscillations overlap. In this case it is
convenient to introduce the partial density of states as shown in
Fig. 7a for zero field. As for the aforementioned FM state of the
array nonstandard behavior of $D(\omega )$ for both branches can
merge caused by long-range character of interaction. As the
previously discussed example, when the discontinuity of the
derivative of density of states has the form $D(\omega )=C_1 +C_2
(\omega -\omega _0 )\Theta (\omega -\omega _0 )$, this corresponds
to the linear dependence of $\omega (\vec {q})$ in the vicinity of
the center of the Brillouin zone, $\omega (\vec {q})-\omega _0
\propto \left| {\vec {q}} \right|$. For the AFM case at various
values of magnetic field this possibility can be realized, but the
situation can be richer compared to the FM case. In particular, this
peculiarity is manifested differently in the density of states for
the upper and lower bands.

\begin{figure}
  \subfigure[ \ h=0.]{\includegraphics[width=\figwidth]{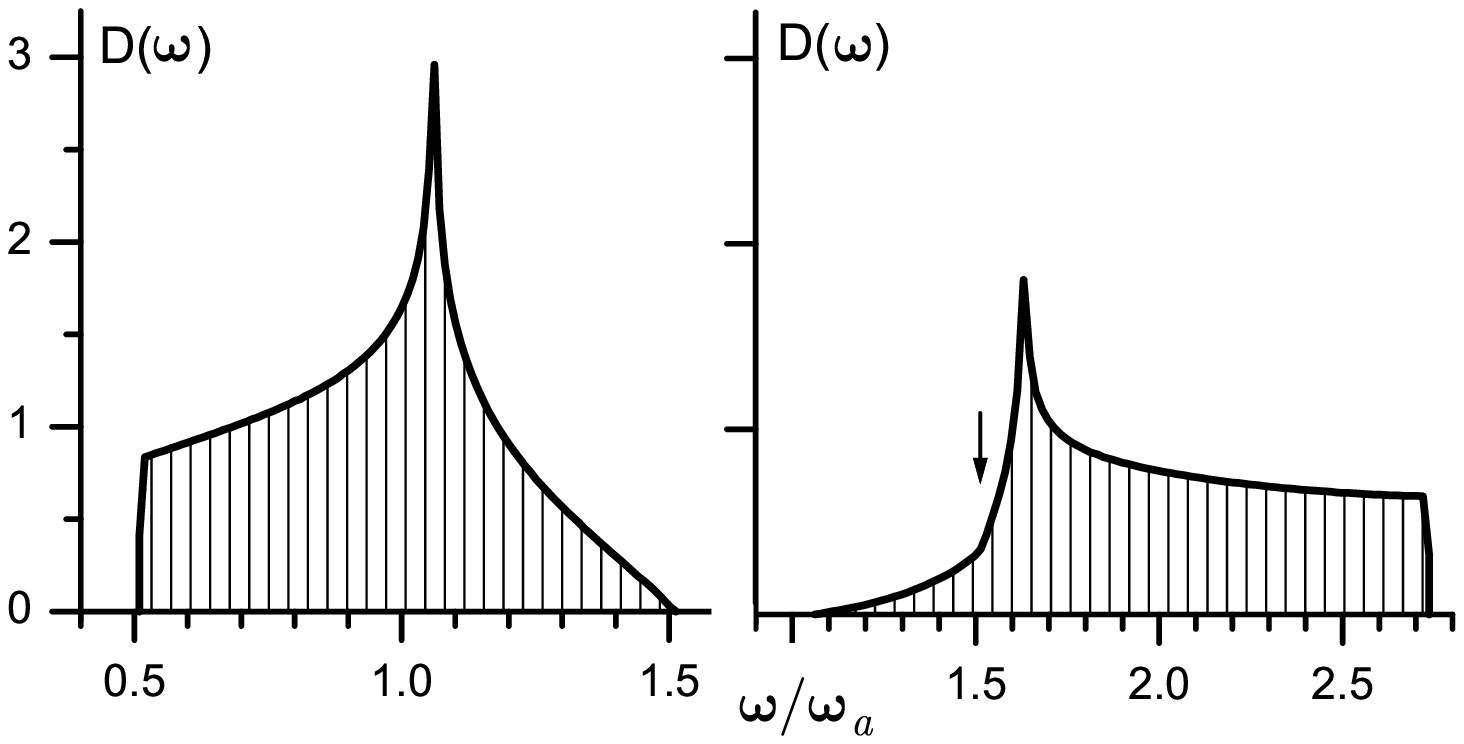}
    \label{exa1}}
    \subfigure[ \ h=2.5.]{\includegraphics[width=\figwidth]{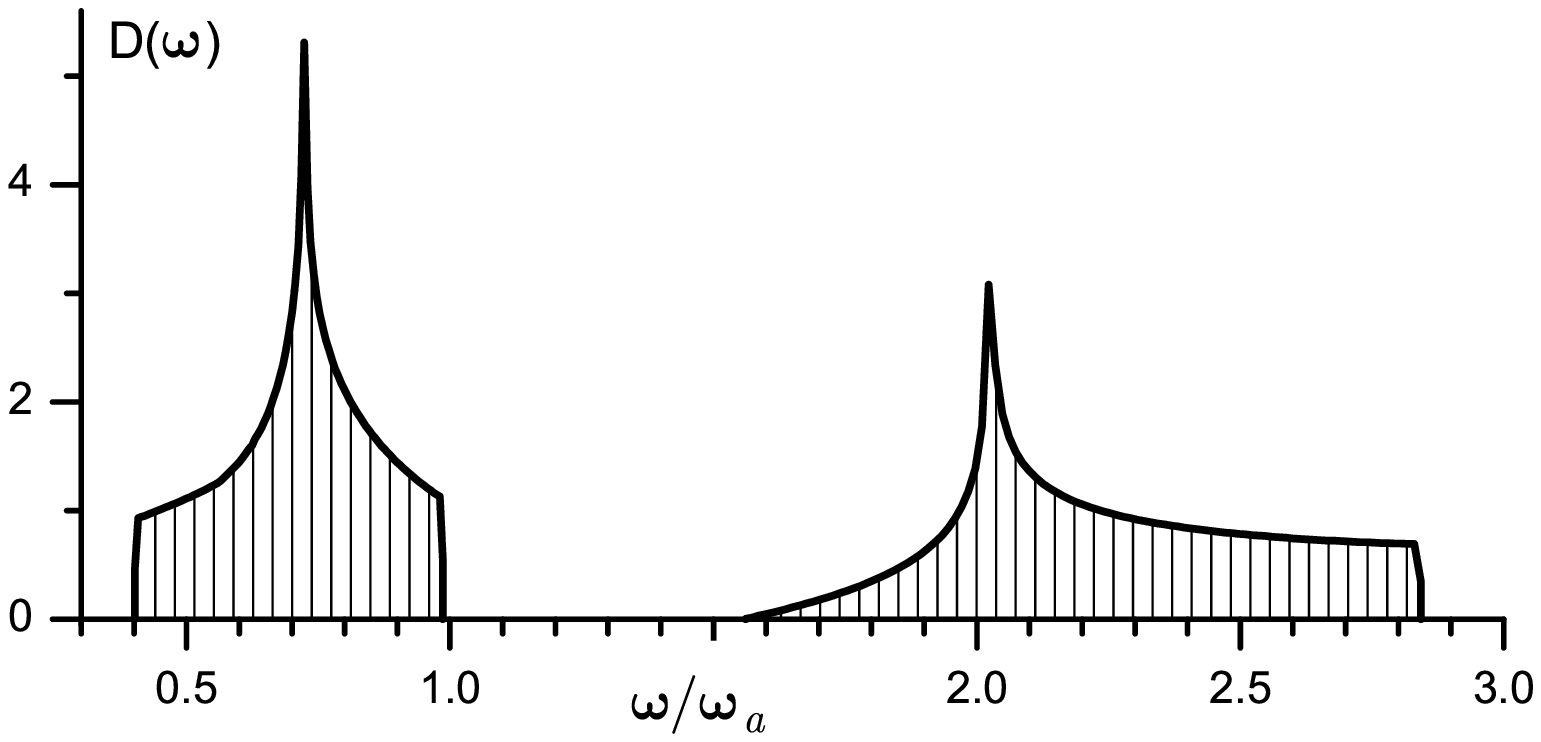}
    \label{exa1}}
    \subfigure[ \ h=5.]{\includegraphics[width=\figwidth]{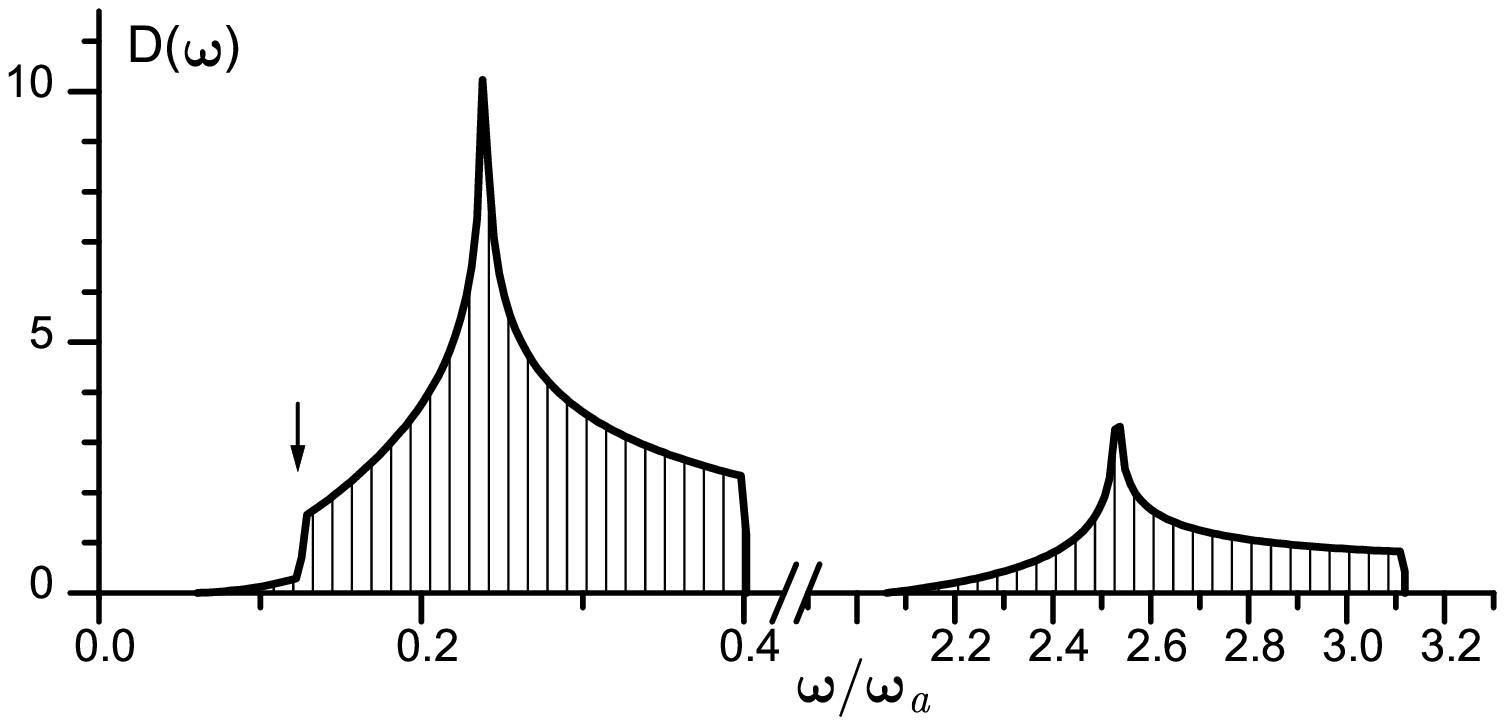}
    \label{exa1}}
       \caption{The density of state for two branches of magnons for the AFM dot
array with $\beta =5$, normalized by the condition $\int {D(\omega
)d\omega } /\omega _a =1$, $\omega _a =\gamma H_a $, for each band,
for different values of the magnetic field. a) the field equals
zero, pay attention that these two branches are overlapped. b)
$h=2.5$, at moderate field values a linear peculiarity of the
density of states is present inside of the lower band of collective
modes; c) $h=5$ presents the case of large fields, where a linear
peculiarity of the density of states corresponds to lower frequency,
and the finite jump is present inside the low energy band. The
characteristic frequencies are mentioned by vertical arrows.
\label{fig7}}
\end{figure}

It happens that the standard scheme of extrema (a minimum in the
centre of the Brillouin zone, maxima on the edges of the Brillouin
zone, saddle points in for symmetrical points like ${\rm X}$(1,0))
take place only for the upper branch and only at large enough field.
In this case in the upper band there are only three Van Hove
singularities (see Fig.~4b,c). For the upper branch the point $\vec
{q}=0$ at any field always defines the global minimum and
peculiarities like $D(\omega )\propto (\omega -\omega _0 )\Theta
(\omega -\omega _0 )$ define the density of states near the lower
edge of the upper band. For all fields the absolute maximum of the
function $\omega _+ (\vec {k})$ having a parabolic dependence is
located at the corners of the Brillouin zone at points such as ${\rm
M}$ (1,1). For this reason, one expects the standard peculiarity of
$D(\omega )$ like a finite discontinuity near the maximal frequency.
For the upper band a nonstandard behavior manifests itself only at
small fields $h<1.2$, when at points like ${\rm X}$(1,0) minima
appear, while saddle points move along directions like (1,0) into
the Brillouin zone. The situation here resembles that described for
the FM case and we do not discuss it here.

The scheme of Van Hove singularities for the lower branch are much less
standard. First of all, for the low band at all values of field the global
maximum is located at the points like ${\rm X}$(1,0). A standard peculiarity
of the density of states like a finite jump corresponds to this frequency.
At $h\ne 0$ (the zero field case is special and will be considered
separately) both at the edges of the Brillouin zone [points like ${\rm
M}$(1,1)] and in its center the minima are present, and the relative depth
of the minima at points $\Gamma $ and ${\rm M}$ is defined by a field value.
Thus, all symmetrical points of the Brillouin zone, $\Gamma $, ${\rm X}$ and
${\rm M}$ are occupied by minima or maxima. For this reason the saddle
points, to which standard logarithmic peculiarities of density of states
correspond, are moved inside of the Brillouin zone along a direction like
(1,0).

The relative position of minima is defined by a value of the
magnetic field. For the lower branch at weak fields ($h<3)$ the
absolute minimum of the function $\omega _- (\vec {k})$ is located
at a point like ${\rm M}$. The parabolic dependence $\omega (\vec
{q})$ corresponds to this minimum, and the latter defines a standard
peculiarity of the density of states like a finite discontinuity
(see Fig.~7a,b). At such values of the field the peculiarity
connected with the point $\Gamma \;(\vec {q}=0)$ lies within the
lower band frequency and manifests itself as a derivative jump
$D(\omega )$ (for example, at $\omega =0.6\omega _a $ for $h=2.5)$
and it is almost invisible in the figures. For higher fields ($h>3$)
the frequency minimum at the point $\Gamma \;(\vec {q}=0)$ becomes
deeper, i.e. the local minimum value of the lower branch of the
spectrum ($\omega _- (\mathrm{M} ))$ lies higher than the frequency
of long-wave oscillations $\omega _- (0)$, (see Fig.~7ñ). In this
case a linear behavior of the density of states $D(\omega )\sim
(\omega -\omega _0 )\Theta (\omega -\omega _0 )$ is clearly seen at
the lower edge of the frequency band for the lower branch. In this
case, within the band there is one more singularity of the density
of states, which is a finite jump of the density of states,
determined by a local parabolic minimum at the point ${\rm
M}\;(1,1)$, see Fig.~7c.

One of the most interesting singularities of the spectrum of the
lower branch is observed in the special case of zero field. First,
in this case the global maxima at the points $\mathrm{X}$ (1,0) have
a non-parabolic dependence $\omega (\vec q)$, $\omega (\vec
q)=\omega_{\max} - c|\Delta \vec q|$, where $\Delta \vec q$ is a
deviation of $\vec q$ from the point $\mathrm{X}$. For this reason,
the upper edge of the energy band is characterized be non-standard
linear Van Hove singularity, $D(\omega )=C(\omega_{\max}-\omega
)\cdot \Theta (\omega_{\max}-\omega )$. On the other hand in the
expansion of $\omega (\vec {q})$ in the center of the Brillouin zone
the linear term in $q$ is absent, and there is a specific saddle
point with four-fold symmetry having a non-analytic behavior, see
Eq.~\eqref{eq21}. However the analysis shows, that this saddle point
leads to the standard logarithmic Van Hove singularity. For small,
but finite value of the external field in the zone center there is a
linear non-analytical behavior of $\omega (\vec {q})\approx \omega
_0 +\alpha \vert \vec {q}\vert $, which is typical for the systems
having dipole interaction. While at the point $\vec q=0$ there is a
minimum and a non-analytic saddle point splits into four standard
saddle points moved towards the points ${\rm X}$.

\section{ Conclusion. }
First let us discuss one possible generalization of our approach for the
problem of interest not considered here. The concrete calculations here have
been done for the most symmetric configurations of the system; namely the
ideal square lattice array, and individual dots having rotational uniaxial
symmetry. The consideration of the same problem with lower symmetry will not
produce any principal difficulties. For a lattice of lower symmetry, such as
rectangular, the same peculiarities for dipolar sums and long wave
asymptotics of dispersion relations are present. The case of dots of lower
symmetry, such as rectangular or elliptic, with almost homogeneous
magnetization within the dots can be easily analyzed, just by adding to Eq.
(\ref{eq1}) extra terms describing more complicated (biaxial) anisotropy of a single
dot. Qualitatively the results will be the same.

In conclusion, the dynamic properties of the array of magnetic dots
of different types without direct exchange interaction between dots
have been investigated. For such systems, only the magnetic dipole
interaction can be a source of dot interaction. Exploiting the
translation symmetry of the array, and by use of the quasiparticle
(magnon) formalism, we have found full spectra for the change of
quasimomentum within all of the Brillouin zone. The direct
measurement of the dependence $\omega (\vec {k})$ can be done by the
Brillouin light scattering method. Our calculations demonstrate that
the dispersion is a strongly increasing function of the dot
thickness. It will be interesting to observe non-monotonic
dependence $\omega (\vec {k})$ and especially, the change of the
character of this dependence with the change of the magnetic field
found here. The most impressive change of the $\omega (\vec {k})$
dependence should appears near the transition from the FM to AFM
states of an array. This property can be of interest for the design
of a new generation of microwave devices in the modern direction of
the applied physics of magnetic nanoparticles, i.e., the so-called
magnonics, which has been widely discussed in the
literature.\cite{magnonics}

An important application of the calculated dispersion laws is the
investigation of the stability of given magnetic states of the
array. It has been found that for arrays with homogeneous
magnetization within a dot directed perpendicular to the array's
plane there will always be a nonzero critical field, either external
or in combination with an anisotropy field. For ferromagnetic state
of array, the instability is developed for the quasimomentum at the
points of the type of (1,0) at the border of the Brillouin zone, and
it leads to a transition to the AFM state with chessboard
orientation of the dot's magnetic moment.

For all magnetic states of the array investigated here, the
dispersion relation is non-analytic as $\vec {k}\to 0$ because of
the long-range nature of the dipolar interaction of oscillating
magnetic moments of dots and the presence of singularities in
dipolar sums. For such modes with finite gap frequency, $\omega
_{0}$ the magnon spectra have the peculiarity of the type $\omega
(\vec {k})-\omega _0 \propto \vert \vec {k}\vert $ as $\vec {k}\to
0$. For the AFM structure without a magnetic field, a new sort of
singularity, leading to saddle point with four-fold symmetry of the
function $\omega ( {\vec {k}})$ near the value $\vec {k}=0$ is
predicted. An important and non-trivial property of most of the
modes considered here is the strong anisotropy of the function
$\omega ( {\vec {k}})$. Note also the decreasing or non-monotonic
dependence of the mode frequency $\omega ( {\vec {k}})$ on the wave
vector $\vec {k}$, observed for non-small interaction of dots.

The non-standard dependence $\omega ( {\vec {k}})$ for small $k$
produce the change of the density of states $D(\omega )$. For the
spectrum of the FM state, instead of standard two-dimensional Van
Hove singularities of a form of finite jump near the gap frequency,
$\omega _0 $, or singularities like $\ln \left( {\omega
_{\mathrm{c}} /\vert \omega _{\mathrm{c}} -\omega \vert } \right)$,
weaker singularities of the form $D(\omega )\sim (\omega -\omega _0
)\Theta (\omega -\omega _0 )$ appear here. On the other hand, for
dispersion relations for the FM state (\ref{eq9}) the change of the
character of its extremum at the point ${\rm X}$ of the type of
(1,0) is predicted at some ratio of the system parameters, $\omega
_0 /\omega _{\mathrm{int}} \simeq 3.6$ (see Fig. 2). For this
critical value of the parameter $\omega _0 /\omega _{\mathrm{int}} $
the density of states has the singularity $D(\omega ) \propto
1/\vert \omega -\omega _{\mathrm{c}} \vert ^{1/4}$, where $\omega
_{\mathrm{c}} $ is the value of frequency at this critical point,
which is stronger than the standard finite jump. The value of the
parameter $\omega _0 / \omega _{\mathrm{int}} $ depends on the
external magnetic field and can be changed continuously during the
experiment.



\section*{Appendix.}
The expressions for collective mode frequencies contain series such
as the dipole sums $\sigma (\vec {k})$ and $\sigma _{c}(\vec {k})$.
Here and below in this Section we will use the dimensionless vector
$\vec {l}$ and the condition $\vec {l} \ne 0$  in the sums is
implied. The mathematical properties of such series are important
not only for this problem, but also for any example of lattice
systems with identical particles coupled by the dipole interaction.

Let us discuss properties of these series. As we will demonstrate
these series have very singular behavior as a functions of
quasimomentum $\vec {k}$, and the double sum converges rather
slowly. This manifests in sum properties near symmetrical points of
the reciprocal lattice $\vec {k}_0 $, especially near the point of
origin $\vec {k}_0 $ = 0 or near the points of the type of
$\mathrm{M}$ (1,1), for which $\vec {k}_0 =\pm (\vec {e}_x \pm \vec
{e}_y )\pi /a$, and $\mathrm{X}$ (1,1), for which $\vec {k}_0 =\pm
(\vec {e}_x)\pi /a$ or $\vec {k}_0 = (\pm \vec {e}_y )\pi /a$.
Analyzing small deviations from these points, $\vec {k}=\vec {k}_0
+\vec {q}$, where $\vec {q}$ is small, one normally has to calculate
derivatives such as [$\partial ^{2}\sigma (\vec {k})$/$\partial
q_{i}\partial q_{j}$] at the point $\vec {k}=\vec {k}_0 $. Term by
term differentiation of the dipole sums gives series like $\sum {
(1/|\vec {l}|)\cdot \exp (i\vec {k}_0 \vec {l}) }$, which are
alternating and converge only conditionally. Moreover, for $\vec
{k}_0 $ = 0, i.e., for the physically most interesting case of long
wave oscillations, the corresponding coefficient of $\vec {q}^{2}$
is described by the divergent series $\sum {1 /| \vec {l} |} $. The
same property is present for the complex sum $\sigma _{c}(\vec
{k})$, which is also important for the description of dipole-coupled
modes. Therefore, the dipole sums at $\vec {k} \simeq 0 $ can be
non-analytical and it results in a non-standard dispersion relation
for oscillations described above.

Let us start with study of the sum $\sigma (\vec {k})$ for small
$\vert \vec {k}\vert $/$k_{B}$. To analyze the behavior of $\sigma
(\vec {k})$ near the point $\vec {k}$ = 0 one can write $\vec {k}=
\vec {q}$, $\vert \vec {q}\vert  \ll  1/a$ and present this series
as
\begin{equation}\label{(A1)}
\sum  {\frac{e^{i\vec {q}\vec {l}}}{\left| {\vec {l}} \right|^3}}
=\sum \frac{1}{\left| \vec {l} \right|^3}\cdot \left[ e^{i\vec
{q}\vec{l}} \cdot e^{-\alpha \vec {l}^2}+e^{i\vec {q}\vec{l}}\cdot
(1-e^{-\alpha \vec {l}^2}) \right]\, ,
\end{equation}
where the multiplier $\exp(-\alpha | {\vec {l}} |^2)$ is chosen to
provide a fast convergence of the corresponding series. The first
term on the right hand side of Eq.~\eqref{(A1)} converges rapidly
for $| {\vec {l}} |>1/\sqrt \alpha $ and at $q =\vert \vec {q}\vert
\to 0$ it contributes as $-q^2{D}'$, where ${D}'=\sum {(1/| {\vec
{l}} | )\exp (-\alpha \vec {l}^2)}$ is an absolutely converging
series. Non-analyticity as $q\to 0$ is defined by the second term,
calculation of which can be simplified. Indeed, at $| {\vec {l}}
|\ll 1/\sqrt \alpha $ it comprises the small multiplier $\alpha \vec
{l}^2$, and the contribution from the region $\left| {\vec {l}}
\right|<1/\sqrt \alpha $ is expected to be small. For the outer
region $| {\vec {l}} |>1/\sqrt \alpha $ one can expect that
discreteness effects are small and the sum can be replaced by the
integral, $\sum_{\vec {l}} \to \int {rdrd\chi } $, where $r$ and
$\chi $ are the polar coordinates. Integration over $\chi $ can be
done exactly through the Bessel function $J_{0 }(qr)$ and the
non-analytical part of the sum $\sigma (\vec {k})$ at small $\vert
\vec {k}\vert $ is determined by the integral $I=2\pi
\int\limits_0^\infty {\left(dr/r^2\right) J_0 (qr)\left(
{1-\exp(-\alpha r^2)} \right)}$, which is convergent both as $r$
$\to $ 0 and $r \to  \infty $. It is convenient to integrate by
parts, which in the case we are interested in ($\alpha \to 0$ )
yields $I=2\pi \int\limits_0^\infty {(dr/r) J_1 (qr)}=2\pi |q |$.
From here we arrive at the expression, used in (\ref{eq6})
$$
\sigma (\vec {k})=\sigma (0) - 2\pi a\vert \vec {k}\vert \
\mathrm{at} \ \vec {k}\to 0 \, .
$$

For the complex sum $\sigma _{c}(\vec {k})$ near the origin $\vec
{k}= \vec {q}$, $\vert \vec {q}\vert \ll 1/a$ the same approach
gives the double integral,
 $$\sigma _{c}(\vec
{q})=\int\limits_0^\infty {\frac{dr}{r^2}} \cdot \int\limits_0^{2\pi
} {d\chi e^{-2i\chi }} \cdot e^{iqr\cos (\chi -\alpha )},$$
 where
$\alpha $ is the angle between $\vec {q}$ and the $x$ axis. Here the
integral over $\chi $ can be done, which gives $\sigma _{c}=2\pi
e^{-2i\alpha } \smallint (dr/r^{2})J_{2}$(\textit{qr}). Further, for
the integral over $r$, which is convergent both as $r \to $ 0 due to
asymptotic $J_{2}(z) \to (1/2)(z/2)^{2}$ as $z\to 0$ and for $r \to
\infty $, it is not necessary to include the regularization
multiplier $\exp(-\alpha r^{2})$ like in \eqref{(A1)}. Finally we
arrive at the following expression
$$\sigma _{c}(\vec {k})=\frac{2\pi }{3}a\frac{(k_x -ik_y )^2}{| {\vec
{k}} |} \ \mathrm{as} \ \vec {k}\to 0.$$

Thus the relation between $\sigma (\vec {k})$ and $\sigma _{c}(\vec
{k})$, $\sigma  (\vec {k})  \to 3 \vert \sigma _{c}(\vec {k})\vert $
as $\vec {k}\to  0$, which is of great importance for the spectral
analysis is asymptotically exact at small values of $\vert \vec
{k}\vert $. This result is in accordance with the numerical
calculation.

Let us discuss now the calculation of the series $\sigma (\vec {k})$
and $\sigma _{c}(\vec {k})$ near the second symmetric point $\vec
{k}_0 $ of the type of (1,1). Here for both series the values of
$D_{ij}$ = [$\partial ^{2}\sigma (\vec {k})$/$\partial k_{i}\partial
k_{j}$]$_{\vec {k}=\vec {k}_0 } $ are described by convergent
alternating series like $\sum  {P_{2p} (\vec {l})} \cdot (-1)^{l_x
+l_y }\left( {l_x^2 +l_y^2 } \right)^{-(2p+1)/2}$, where
$P_{2p}(\vec {l})$ is the polynomial in $l_{x}, \, l_{y}$ of degree
$2p$, $p= 1 $ and $p = 2$ for components $D_{ij}$ in the case of
$\sigma (\vec {k})$ and $\sigma _{ c}(\vec {k})$, respectively.

By using symmetry relations, one can demonstrate that the value of
the function $\sigma (\vec {k})$ at $\vec {k}$ of type (1,1) is
described only by one sum, $D = -\sum {(-1)^{l_x +l_y }\left( {l_x^2
+l_y^2 } \right)^{-1/2}} $, the value of which can be easily found
numerically, $D \cong 1.61554$ . As a result the following
expression is obtained
$$\sigma (\vec {k}) \cong  \sigma (\vec {k}_0 )+
\frac{1}{4} Da^{2}\vert \vec {k}-\vec {k}_0 \vert ^{2}.$$

For analysis of the sum $\sigma _{c}(\vec {k})$ near the point $\vec
{k}=\vec {k}_0 $ we note that because of the evident relations
$$\sum_{\vec {l}\ne 0} (l_x^2 -l_y^2)\cdot ( l_x^2 +l_y^2
)^{-5/2}(-1)^{l_x +l_y }= 0 ,$$
 the value of $\sigma _{c}(\vec {k}_0
)= 0$. The next terms of the expansion $\sigma _{c}(\vec {k})$ over
the vector components $\vec {q}= \vec {k} - \vec {k}_0 $ can be
written easily,
\begin{multline}
\sigma _{c}(\vec {k}) = 2iq_{x}q_{y} \sum {\frac{l_x^2 l_y^2
}{\left( {l_x^2 +l_y^2 } \right)^{5/2}}(-1)^{l_x +l_y }} + \\
+\frac{1}{2}\left( {q_x^2 -q_y^2 } \right)\sum {\frac{l_y^2 (l_x^2
-l_y^2 )}{\left( {l_x^2 +l_y^2 } \right)^{5/2}}(-1)^{l_x +l_y }} .
\end{multline}

They comprise both real terms proportional to $q_x^2 -q_y^2 $ and
imaginary terms like $q_xq_y$. After simple algebra, these terms are
presented via the sum $D$ introduced above and one more convergent
sum of similar structure
 $$D_{1}=-\sum
(l_x^2 -l_y^2 )^2(-1)^{l_x +l_y }/\left( {l_x^2 +l_y^2 }
\right)^{-5/2} \cong  3.31891.$$
 Finally, in the quadratic
approximation over $\vec {q}= \vec {k} - \vec {k}_0 $ near the point
$\vec {k}_0 $ of the type (1,1) one can present the sum as
\begin{equation}\nonumber
\sigma _{c}(\vec {k}) =\sigma _{c}(\vec {k}_0 )+i(D_{1 }- D) a^{
2}q_{x}q_{y} + \frac{1}{4}D_{1} a^{ 2}(q_x^2 -q_y^2 ).
\end{equation}

This function, in contrast with $\sigma (\vec {k})$ near the same
point $\vec {k}_0 $, is anisotropic. However, expressions for the
frequencies contain $\sigma (\vec {k})$ and $\sigma _{c}(\vec {k})$
only in the combination $\sigma ^{2}(\vec {k})-9\vert \sigma
_{c}(\vec {k})\vert ^{2}$, therefore for presentation of the
spectrum of collective modes accurate to $\vert \vec {q}\vert ^{2}$
it is not essential to take $\sigma _{c}(\vec {k})$ into account,
and to the accuracy of $q^{2}$ the spectrum near the point $\vec
{k}_0 $ of the type (1,1) is radially symmetric over components of
the vector $\vec {q}$ .

The point with $\vec {k}_0 $ of the type (1,0) including the points
like $\vec {k}_0 =\pm \pi \vec {e}_x /a$ or $\vec {k}_0 =\pm \pi
\vec {e}_y /a$, possess lower (biaxial) symmetry than the points
$\vec {k}= 0$ or points of the type (1,1). The rather complicated
study in the vicinity of this point ($\vec {q}= \vec {k} - \vec
{k}_0 $ is small) shows the analytical dependence for the components
of $\vec {q}$, in this case for the components of $\vec {q}$
parallel and perpendicular to $\vec {k}_0 $
 $$\sigma (\vec {k})- \sigma (\vec {k}_0 )=\frac{1}{2}a^{2} (D_{\vert
\vert } q_{\vert \vert }^{2} -D_{\bot }q_{\bot }^{2}),$$
  where $D_{\vert \vert }$ and $D_{\bot }$ are convergent series of the form
\begin{multline}
D_{\vert \vert }=-\sum{\frac{ l_x^2 (-1)^{l_x
}}{\left( {l_x^2 +l_y^2 } \right)^{5/2}}  }\cong  0.6354\,, \\
D_{\bot }=\sum\limits_{\vec {l}\ne 0}{\frac{ l_y^2 (-1)^{l_x
}}{\left( {l_x^2 +l_y^2 } \right)^{5/2}}  }\cong 1.256 \,.
\end{multline}

\end{document}